\documentclass[aps,prb,preprint,superscriptaddress,amsmath,footinbib]{revtex4-1}
\usepackage{graphicx}
\usepackage{bbold}
\usepackage{hyperref}
\usepackage{import}
\usepackage{color}
\usepackage{soul}

\newcommand{\ket}[1]{\ensuremath{\,|#1\rangle}}

\begin{document}


\title{Quantum information transfer using photons} 



\author{T. E. Northup}
\email[Correspondence and requests for materials should be addressed to ]{tracy.northup@uibk.ac.at.}
\affiliation{Institut f{\"u}r Experimentalphysik, Universit{\"a}t Innsbruck, Technikerstra{\ss}e 25, 6020 Innsbruck, Austria}
\author{R. Blatt}
\affiliation{Institut f{\"u}r Experimentalphysik, Universit{\"a}t Innsbruck, Technikerstra{\ss}e 25, 6020 Innsbruck, Austria}
\affiliation{Institut f\"ur Quantenoptik und Quanteninformation der \"Osterreichischen Akademie der Wissenschaften,
Technikerstra{\ss}e 21a, 6020 Innsbruck, Austria}
\date{\today}

\begin{abstract}

Optical communication channels have redefined the purview and applications of classical computing; similarly, photonic transfer of quantum information promises to open new horizons for quantum computing.
The implementation of light-matter interfaces that preserve quantum information is technologically challenging, but key building blocks for such devices have recently been demonstrated in several research groups.
Here, we outline the theoretical framework for information transfer between nodes of a quantum network, review the current experimental state of the art, and discuss the prospects for hybrid systems currently in development.

 \end{abstract}


\maketitle 

%
%


Quantum physics is one of the most important intellectual achievements of the 20th century. It has profoundly changed our view of the world and offered revolutionizing technologies such as the laser and NMR imagery. During the last two decades, quantum physics has entered the fields of information processing and communication. New quantum concepts for computational purposes have stimulated enormous efforts to develop platforms and protocols that enhance classical computation and communication devices. While quantum information processors rely on the manipulation of quantum bits (qubits), quantum communication naturally seeks to operate with photons. Thus, the transfer of quantum information between two stationary nodes with photons --- often dubbed flying qubits --- becomes a focal point of the new quantum information technology.

Stationary qubits such as atoms, quantum dots, and superconducting circuits are able to store quantum information for certain times, known as coherence times, which are limited by their respective coupling to the surrounding environment. Photons, on the other hand, have very little interaction with one another and are ideally suited to transport information over large distances. Quantum information transfer with photons is thus understood as remote distribution of information that preserves the underlying quantum states. Quantum communication \cite{Gisin07} has already matured to encompass practical applications such as quantum key distribution and cryptography, and visionary concepts such as the ``quantum internet" \cite{Kimble08a} have been conceived and are currently being pursued in the laboratory.

With such technology at hand, new areas of physics become accessible: the transfer of quantum information allows one to study foundations of quantum physics such as non-locality issues, i.e., non-classical states spanning classical distance scales and the quantum correlations of such extended systems. With the development of quantum processors and interfaces between stationary and flying qubits, even distributed quantum computation appears possible, promising scalable quantum information processing at multiple sites on the same device or across longer distances within a network.  
The transfer of quantum information with photons thus bridges the microscopic and macroscopic worlds and may pave the way toward new applications of quantum physics in our everyday life.

\section{Requirements for quantum information transfer}

Quantum information is typically stored in superpositions $|\Psi\rangle = c_0|0\rangle+c_1|1\rangle$ of two-level systems $\{|0\rangle,|1\rangle\}$, which can be realized as long-lived electronic states in atomic and solid-state quantum bits. The first requirement for quantum information transfer (QIT) is thus a quantum node, at which quantum information is not only stored but also generated and processed.  In this review, both quantum nodes and quantum memories are discussed.  In contrast to quantum nodes, quantum memories are intended solely for information storage; for example, quantum information generated at one or several remote nodes may be cached in a memory before it is processed.

Next, in order to transmit quantum information, a quantum channel is required.  Long-distance quantum communication, earthbound or even involving satellites, can be realized by free-space optical channels \cite{Yin12,Ma12}, but it is often convenient to take advantage of optical fibers for photon transport.  Finally, linking quantum nodes and quantum channels requires the implementation of a light-matter quantum interface. The technical implementation of the qubits dictates the design and construction of the interfacing element, whereas the optical quantum channel should be compatible with fiber-based or free-space optical technology available for classical communication. High-fidelity qubits have already been realized in diverse experimental settings, and the use of photons for quantum communication and quantum key distribution is well established \cite{Gisin02,Gisin07}. 
The crucial element required for QIT using photons is thus the light-matter interface that allows one to map (stationary) atomic or solid-state qubits to (flying) photonic qubits.

This review therefore focuses on the realization of such quantum interfaces, and in particular, emphasizes recent experimental progress in this field since the review by Kimble \cite{Kimble08a}. The paper is organized as follows: In the first section, we describe the general function of a light-matter interface, distinguish between deterministic and heralded protocols, and highlight the role of optical cavities.  The next section gives a more detailed description of several key experiments for remote atom--atom entanglement, mediated by light.
Subsequently, we outline how quantum repeaters enable long-distance quantum communication. The final section highlights upcoming techniques, possible improvements and perspectives for future developments.

\section{Blueprints for a light-matter interface}

\begin{figure}
 \includegraphics[width=\textwidth]{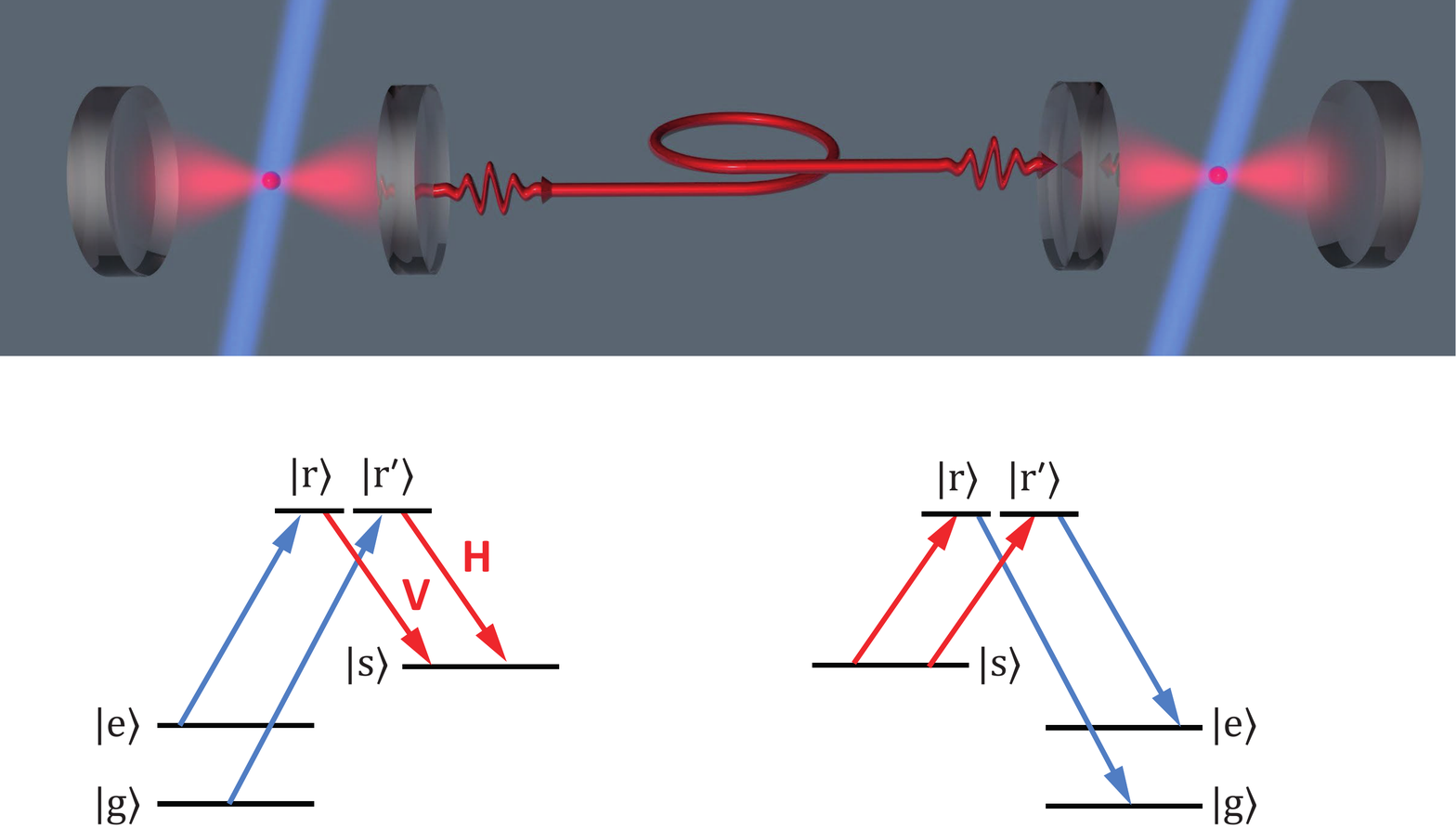}
 \caption{\label{fig1} A modified version of the deterministic atom-photon interface proposed by Cirac, Zoller, Kimble, and Mabuchi \cite{Cirac97}.  Single atoms in optical cavities are linked by a photonic channel.  Information stored in two electronic ground states of the first atom is mapped onto a cavity photon (red) via classical laser fields (blue), which couple both ground states to the same final state.  The photon exits the first cavity and travels down the channel.  It is then mapped onto the atom in the second cavity via the time-reversed laser fields.}
 \end{figure}

The first blueprint for a light-matter interface was conceived just a few years after Shor's factoring algorithm sparked widespread interest in quantum computing \cite{Shor94} and strings of trapped ions were shown to offer a promising architecture \cite{CiracZoller95}.
In their seminal proposal \cite{Cirac97}, Cirac and colleagues suggested that such atom-based computers could be linked together by transferring information to and from photons via an optical resonator.  
By tuning the resonator's frequency near an atomic transition, one takes advantage of the dipole coupling between the atoms and the cavity field, that is, the fundamental interaction of cavity quantum electrodynamics \cite{Haroche06}.  
This interaction allows a single quantum of information stored in the electronic states of an atom to be reversibly exchanged with a single photon in the field.
A schematic representation of such an interface is shown in Fig. \ref{fig1}, with details presented in Box 1.
Note that while the original proposal encoded quantum states in the photon number basis \{\ket{0}, \ket{1}\}, here we use a basis of linearly polarized photons \{\ket{H}, \ket{V}\}, which has the advantage that it is more robust to losses in the transmission channel.  A third option would be to use a time-bin qubit, in which the photon exits the cavity in a superposition of two possible time windows \cite{Brendel99}.

This scheme enables the \textit{deterministic} transfer of a quantum state between remote quantum nodes.  
An experimental implementation presents several challenges: neutral atoms or ions must be stably trapped and positioned within an optical cavity, which should have low scattering and absorption losses and a high atom-cavity coupling rate with respect to decoherence rates of the system.
Even using a state-of-the-art cavity that satisfies the strong coupling criterion \cite{Kimble08a}, the transfer process will nevertheless accumulate errors, requiring quantum error correction \cite{vanEnk97,vanEnk97a}.

More recently, a second framework for a quantum interface has emerged that promises a reduced technical overhead 
\cite{Duan10,Monroe13}.
This implementation is \textit{heralded} rather than deterministic, that is, not every attempt to transfer quantum information is successful, but the successful cases are flagged.  
The user thus executes the transfer protocol repeatedly until the heralding signal is received.
It may seem at first glance that heralded transfer must be less efficient than deterministic transfer, but in fact, that depends on the physical parameters of the implementation.
Furthermore, the heralding process is robust to certain errors, 
so that the quantum state is transferred faithfully.

\noindent
\fbox{
  \parbox{\textwidth}{
\textbf{Box 1: A deterministic atom-photon interface} \\ 
In a scheme based on Ref. \onlinecite{Cirac97} and illustrated in Fig. \ref{fig1}, at any given time, a single trapped atom interacts with the cavity field. 
That atom may be part of a larger ensemble, i.e., a small-scale quantum computer.  (The interaction of the other atoms with the cavity may be turned off via electronic shelving to an uncoupled state, or by positioning the atoms so that they do not couple to the spatial mode of the field.)  An atom is prepared in one of two long-lived states,  $\ket{g}$ and $\ket{e}$, or in a superposition of both.  The cavity is nearly resonant with the transitions between a third ground state, $\ket{s}$, and the excited states $\ket{r}$ and $\ket{r'}$.  If a classical field $\Omega_1(t)$ coupling $\ket{g}$ to $\ket{r}$ is applied to the atom, and if the frequency difference between $\Omega_1(t)$ and the cavity matches the energy gap between $\ket{g}$ and $\ket{s}$, then a Raman process transfers an atom in $\ket{g}$ to $\ket{s}$, coherently generating a single photon in the cavity.  Similarly, a field $\Omega_2(t)$ 
generates a single photon if the atom starts in $\ket{e}$, again mapping the atom to $\ket{s}$.  For the appropriate choice of electronic states, the photons generated by $\Omega_1(t)$ and $\Omega_2(t)$ are orthogonally polarized; let us assume that their polarization is either horizontal ($\ket{H}$) or vertical ($\ket{V}$).  The simultaneous application of $\Omega_1(t)$ and $\Omega_2(t)$ then implements the transfer of a quantum state from an atom to a photon:
\begin{align*}
\sin{\theta}\ket{g} + e^{i\varphi}\cos{\theta}\ket{e} \rightarrow \sin{\theta}\ket{H} + e^{i\varphi}\cos{\theta}\ket{V},
\end{align*}
where $\theta$ and $\varphi$ parameterize the quantum state on the Bloch sphere, or in the case of the photon, the Poincar\'e sphere.
The photon exits the cavity in a well-defined spatial mode, and its temporal shape can be symmetrized by tailoring the pulse shapes of $\Omega_1(t)$ and $\Omega_2(t)$ \cite{Kuhn99,Keller04}. 
The photon is then sent over an optical fiber link to a second cavity, where the process is time-reversed: a second atom is initialized in $\ket{s}$, and as the photon enters the cavity, the mirror-image waveforms $\Omega_1'(t)$ and $\Omega_2'(t)$ map the atom to $\ket{g}$ and $\ket{e}$.
}}

\begin{figure}
 \includegraphics[width=\textwidth]{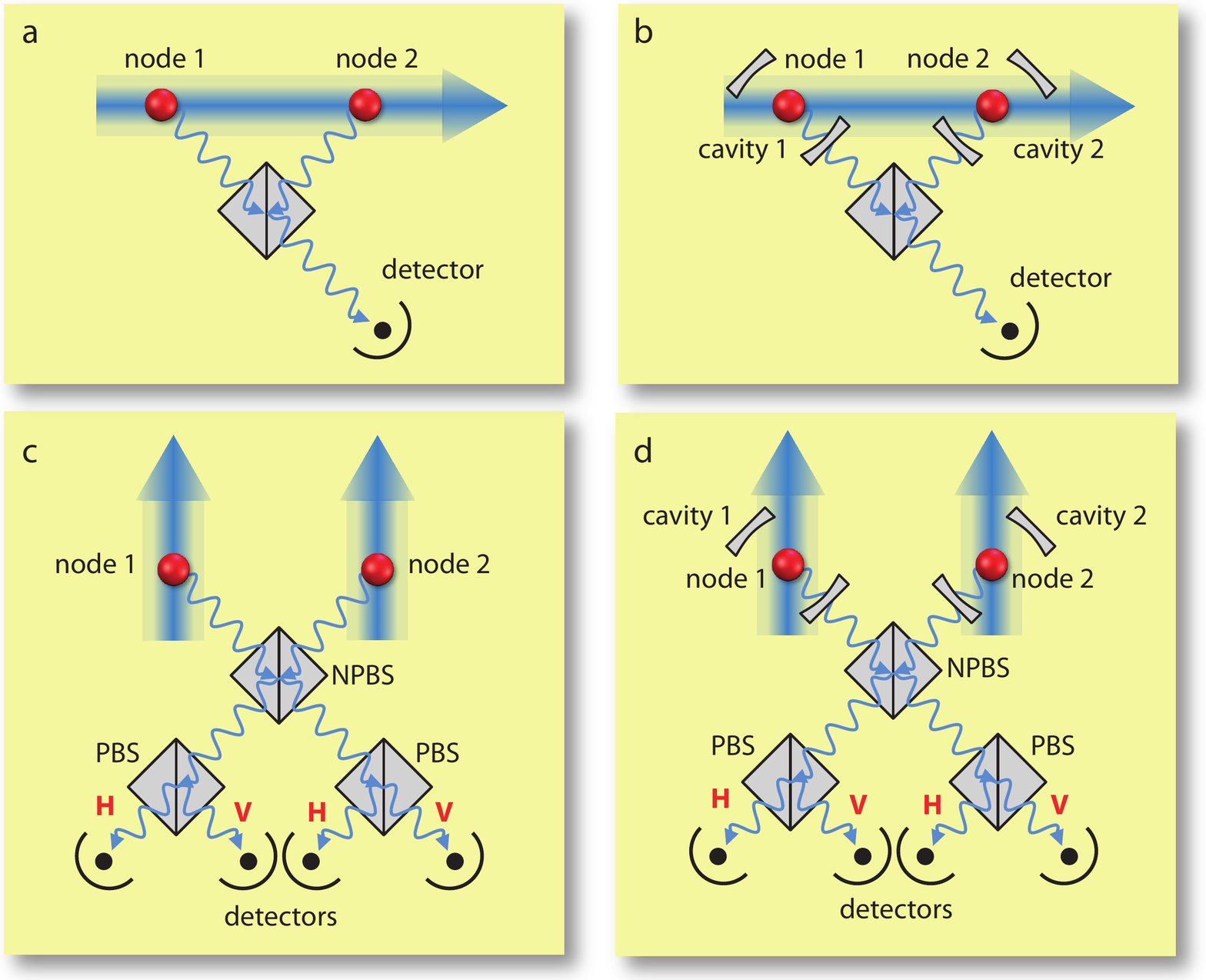}
 \caption{\label{fig2} A heralded interface between two quantum memories does not rely on a direct optical connection between the memories.  \textbf{a}  In one implementation, both memories are weakly excited by a laser pulse.  There is a small probability that the pulse maps each memory from an initial to a final quantum state, generating a single photon in the process.  The photon paths from the memories are then combined at a beamsplitter.  If a detector at the beamsplitter output registers a single photon, the indistinguishability of the two paths projects the memories into an entangled quantum state. \textbf{b} The memories can be placed within optical cavities to enhance the photon-collection probability.  \textbf{c}  A second method requires entangling each memory with the polarization state of a photon.   Again, a beamsplitter (NBPS) removes `which-way' information about the photon paths.  The simultaneous measurement of orthogonally polarized photons H and V at the two polarizing beamsplitter (PBS) outputs generates entanglement. \textbf{d} Here as well, optical cavities increase the rate at which remote entanglement is generated.}
\end{figure}

Heralded transfer can be seen as a three-step process, comprising photon generation, measurement, and teleportation (Fig. \ref{fig2}).  
The first step is either to weakly excite each of two spatially separated atoms so that at most one of them generates a photon \cite{Cabrillo99} or to entangle each atom with a photon \cite{Feng03, Duan03, Simon03}.
Next, the photon paths from the two atoms coalesce at a beamsplitter,  which erases `which-way' information. 
If a certain detection outcome is recorded on photon counters at the beamsplitter outputs, the measurement event projects the remote atoms into an entangled state.
Finally, this atom-atom entanglement is used as a resource to teleport quantum information between the two sites \cite{Bennett93}.  
For example, assume that the quantum information is encoded in a third atom, a neighbor of the second.  A joint measurement on the second and third atom is performed, and the result is sent over a classical channel to the distant first atom.  A rotation conditioned on this result transfers the state of the third atom onto the first. 

\section{Experimental building blocks}

In the past decade, a handful of research groups have implemented both deterministic and heralded schemes and are working to extend those results across a range of experimental settings.  In the deterministic case, a necessary precursor has been the development of sophisticated techniques for trapping and manipulating atoms and ions within low-loss resonators.  Neutral atoms can now be confined to the cavity region by an off-resonant standing-wave field \cite{Ye99,Puppe07} or by a transverse field \cite{Sauer04a,Nussmann05,Khudaverdyan09} for times on the order of a minute \cite{Ritter12}.  Furthermore, the transverse field can be used as a conveyor belt  \cite{Schrader01} to position the atoms precisely with respect to the cavity mode \cite{Sauer04a,Nussmann05,Khudaverdyan09}.  In experiments in the strong-coupling regime, the coherent interaction between a single atom and a photon dominates the system dynamics \cite{Miller05}.

Using trapped ions, it is comparatively easier to obtain long storage times, since a Paul trap for charged particles is typically several orders of magnitude deeper than an optical-dipole trap \cite{Paul90}.  However, in order to obtain a high atom-cavity coupling rate, a small cavity mode volume is necessary.   Such a small mode volume is difficult to achieve in an ion-trap setting while maintaining optical access for lasers, 
and while avoiding perturbations of the trapping potential due to the dielectric cavity mirrors.  
Thus, experiments have not yet reached the single-ion strong-coupling regime, although fiber-based cavities offer a promising route \cite{Steiner13}.  Transport and positioning of the ion within the cavity mode can be achieved by translating the cavity via piezo stages with respect to the Paul trap \cite{Guthoehrlein01,Russo09}.  

Both neutral atoms and ions can be localized to length scales much shorter than the cavity standing wave via cooling to the vibrational ground state \cite{Boozer06,Stute12a,Reiserer13}, a technique first implemented in ion traps \cite{Leibfried03}.  Cooling has also been demonstrated using the cavity to extract blue-detuned photons from the system \cite{Horak97,Chan03,Maunz04,Fortier07,Leibrandt09} and via feedback to the dipole field seen by intracavity atoms \cite{Kubanek09,Koch10}.

With these techniques in hand, deterministic transfer has been demonstrated in both directions: from light onto matter and from matter onto light.  Using a weak coherent state, that is, a laser pulse with a mean photon number of about one, it was shown that this state could be reversibly transferred to and from the hyperfine levels of a trapped cesium atom in a cavity \cite{Boozer07a}.  In this case, information was encoded in the number basis \{\ket{0}, \ket{1}\}, representing the absence or presence of a photon.  Given a lossy transmission channel, however, information encoded in $\ket{1}$ at the input may be identified as $\ket{0}$ at the output, thus reducing the fidelity of the transfer process.  An important step was thus the realization of a polarization-based interface, linking two atomic hyperfine states with orthogonal cavity photons \cite{Wilk07b}.  Assuming both polarizations experience equal losses in the optical channel, the encoding is robust in the sense that losses do not affect the process fidelity, although they reduce its efficiency.  Polarization states of light can be mapped into and out of an atom-cavity system, with coherence times exceeding 100 $\mu$s \cite{Specht11}.

This light-matter interface has now been extended to spatially separated systems, linked by optical fiber.  A single trapped rubidium atom was entangled with a cavity photon, which was sent over optical fiber to a second laboratory and stored in a Bose-Einstein condensate, thus generating remote entanglement \cite{Lettner11}.  More recently, quantum information was transferred from one atom to a cavity photon, then mapped to a second atom in a distant cavity \cite{Ritter12}.  These results not only synthesize key techniques for a quantum network, but also highlight the contrast between quantum nodes --- exemplified by atoms in cavities --- and quantum memories, where quantum-degenerate gases offer long storage times. State transfer from atom to photon has also been demonstrated in an interface based on a trapped calcium ion \cite{Stute13} (Fig. 3).  The initial quantum states of ions can be prepared deterministically, and because techniques for coherent manipulation and detection of ions are well established, the final states can be read out from the ions in an arbitrary basis  \cite{Leibfried03,Haeffner08}.

What does it mean to say that the transfer process in these experiments is deterministic?  The key concept is that the state of an atom is mapped onto a cavity photon, and vice versa, with a probability approaching one.  This deterministic character is due to the fact that the dipole interaction between atom and cavity is coherent and thus reversible.  The probability is not exactly one because there is a small chance for the atom to spontaneously emit a photon into free space during this process, which erases information from the system.  One should also note that even if a cavity photon is created deterministically, it can be scattered or absorbed in the cavity mirrors or, after it exits the cavity, in an optical channel.  For example, state transfer has been demonstrated from atoms to photons with a probability of $16\%$\cite{Stute13} and from photons to atoms with a probability of $20\%$\cite{Ritter12}, limited by mirror losses and by the cavity coupling strength.

Heralded schemes, in contrast, do not need to work every time and are thus suited to a wider range of experimental systems.  In fact, light-matter entanglement, a key building block, has been realized in diverse systems including ions \cite{Blinov04,Olmschenk09}, single atoms \cite{Volz06}, atomic ensembles \cite{Matsukevich05,Sherson06}, nitrogen-vacancy (NV) centers in diamond \cite{Togan10}, diamond crystals \cite{Lee11}, and quantum dots \cite{Gao12,Degreve12,Schaibley13}.  In these schemes, optical cavities are no longer a necessary ingredient, but they greatly enhance the photon collection rate and also provide the ability to tune the entangled state parameters \cite{Wilk07b,Stute12,Ritter12}.

 \begin{figure}
 \includegraphics[width=0.7\textwidth]{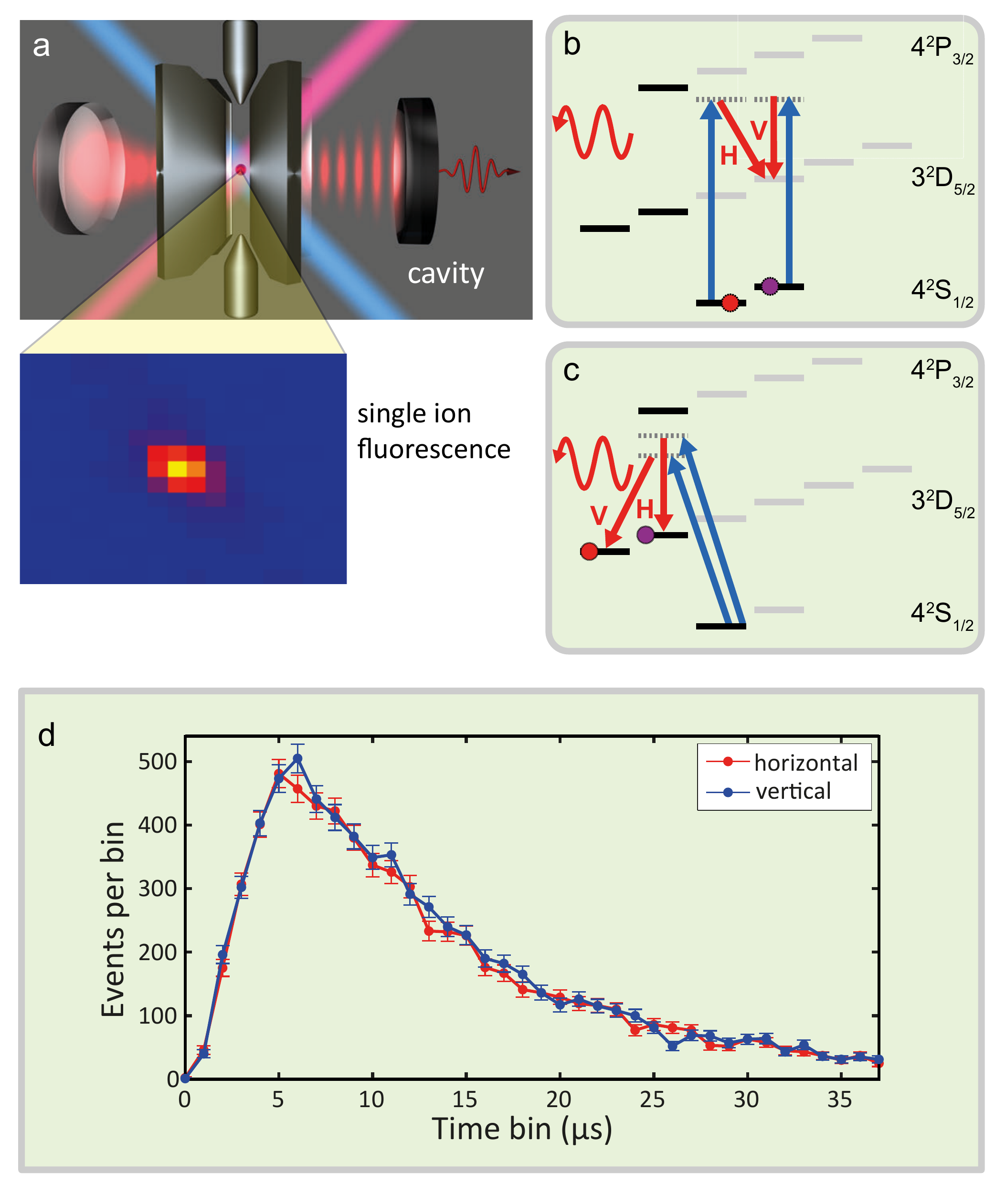}
 \caption{\label{fig3} One example of a light-matter interface.  \textbf{a} A linear Paul trap confines single $^{40}\text{Ca}^+$ ions within a near-concentric optical cavity.  Piezo stages translate the cavity with respect to the trap, so that a single ion couples to an antinode of the cavity field. \textbf{b} A deterministic interface: quantum information stored in two ground states of the $4^2S_{1/2}$ manifold (split by a magnetic field) is mapped to the polarization of the cavity photon via a bichromatic Raman field detuned from the $4^2S_{1/2} - 4^2P_{3/2}$ transition at 393~nm.  The photon is near-resonant with the $3^2D_{5/2} - 4^2P_{3/2}$ transition at 854~nm. Schematic adapted from Ref. \onlinecite{Stute13}. \textbf{c}  The ion-cavity system can also be used as one node of a heralded interface.  The ion is prepared in one ground state, and ion-photon entanglement results from coherently coupling the ion to two states in the $3^2D_{5/2}$ manifold.  \textbf{d}  The two couplings generate orthogonal horizontally (H) and vertically (V) polarized photons with equal probability.  To create remote ion-ion entanglement, the photon would then be combined at a beamsplitter with a second photon from a distant node.  Schematics adapted from Ref. \onlinecite{Stute12}. } 
 \end{figure}

Light-matter entanglement is not in itself sufficient to achieve heralded entanglement between remote quantum nodes.  In addition, the photons sent from both nodes to a common location must be indistinguishable, so that each photon carries no information that could reveal its origin.  Such indistinguishability is typically verified by the observation of Hong-Ou-Mandel interference, in which two identical photons impinging on a beamsplitter always exit as a pair at one output \cite{Hong87}.  
When two quantum memories are weakly excited in order to entangle remote nodes \cite{Cabrillo99}, this is equivalent to entangling the electronic state of each memory with the photon number state.
In this case, not only photon indistinguishability but also interferometric path stability between the nodes is required.  Heralded remote entanglement has been demonstrated between atomic ensembles \cite{Chou05}, single ions \cite{Moehring07}, neutral atoms \cite{Hofmann12}, crystals doped with rare-earth ions \cite{Usmani12}, and NV centers \cite{Bernien13}.  


The final step in heralded information transfer, teleportation, requires a local measurement, the result of which is transmitted over a classical channel.  For example, in Ref. \onlinecite{Olmschenk09}, an initial state $\alpha|0\rangle_A + \beta|1\rangle_A$ was stored in the hyperfine states $|0\rangle$ and $|1\rangle$ of a trapped Yb$^+$ ion at site A.  Given a second ion at B, one meter distant, the ion-photon entangled states
\begin{align*}
\Psi_A &= \alpha |0\rangle_A|\nu_{\text{blue}}\rangle_A  + \beta |1\rangle_A|\nu_{\text{red}}\rangle_A \\
\Psi_B &= |0\rangle_B|\nu_{\text{blue}}\rangle_B  +|1\rangle_B|\nu_{\text{red}}\rangle_B 
\end{align*}
were created, then projected by photon detection into the ion-ion entangled state
$\alpha |0\rangle_A |1\rangle_B  - \beta |1\rangle_A |0\rangle_B$,
where $\nu_{\text{blue}}$ and $\nu_{\text{red}}$ are photon frequencies.
The final step was then a local rotation of the ion at A, followed by a fluorescence measurement projecting it into either $|0\rangle$ or $|1\rangle$.  This classical result was sent to B and determined the appropriate rotation to prepare $\alpha|0\rangle_B + \beta|1\rangle_B$.
This demonstration of teleportation between remote ions followed
earlier results using ions stored in the same trap \cite{Riebe04,Barrett04}; 
with the the same system of remote Yb$^+$ ions, a heralded quantum gate was also implemented \cite{Maunz09}.
Recently, teleportation has been achieved with both neutral atoms \cite{Noelleke13} and atomic ensembles \cite{Bao12,Krauter13}.




In order to label either a deterministic or a heralded implementation as ``quantum," one must show that the result surpasses the best possible classical outcome.  Moreover, even if a process satisfies this criterion, it is interesting to quantify the fidelity of the process: how well does the transferred state reproduce the initial state?  This fidelity can then be compared with theoretical bounds in order to understand whether it would be possible to implement error correction protocols, an essential component of a scalable quantum network.  

\begin{figure}
\includegraphics[width=0.33\textwidth]{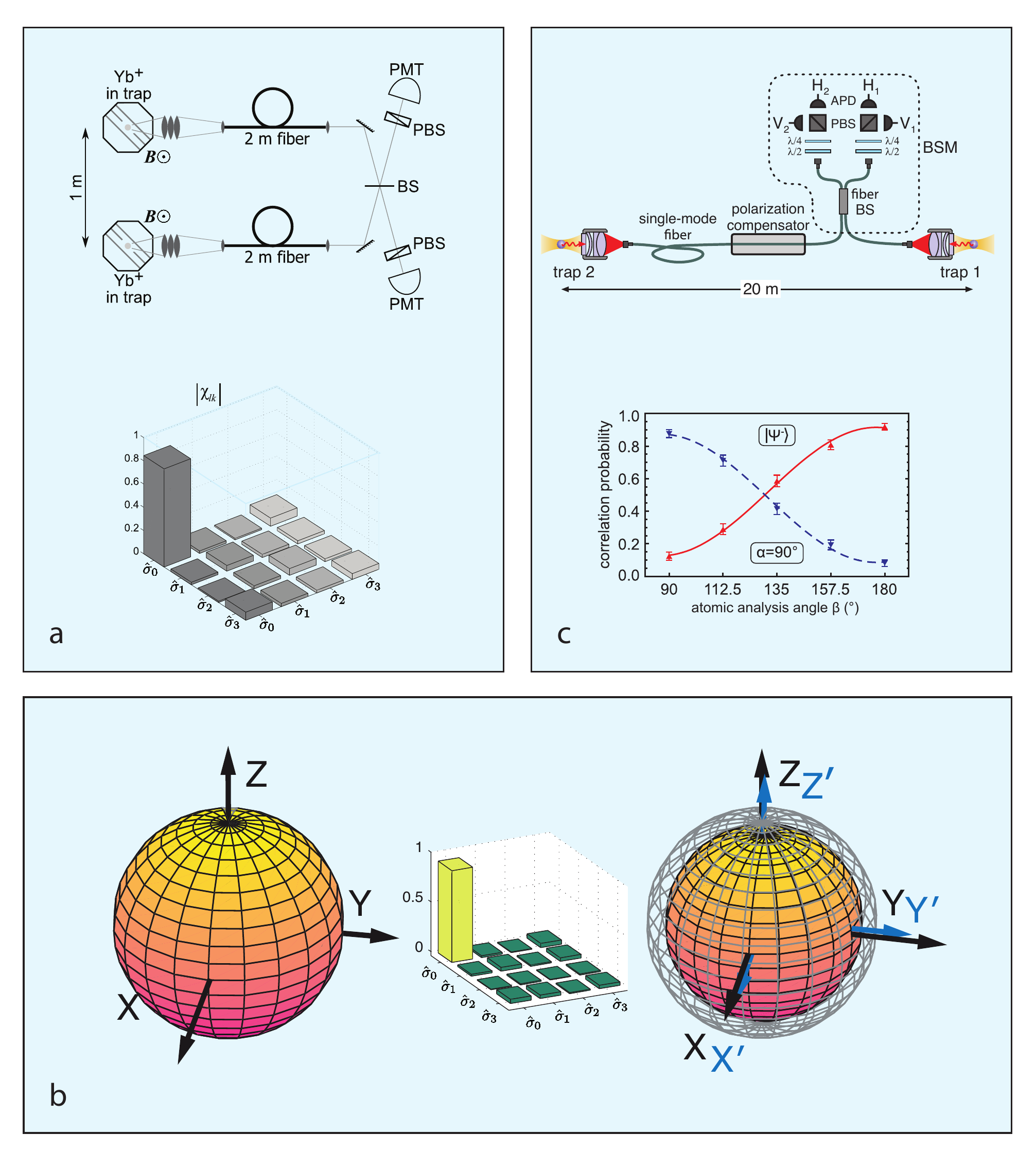}
 \caption{\label{fig4} A quantum process described by the matrix $\chi$ maps an arbitrary input density matrix $\rho_{\text{in}}$ to the output matrix $\rho_{\text{out}}$.  Since the transfer process should leave $\rho_{\text{in}}$ unchanged, in the ideal case all entries of $\chi$ would be zero except the identity term $F = \chi_{00}$.
\textbf{a} Process tomography is used to characterize teleportation of quantum information between two Yb$^{+}$ ions, separated by one meter \cite{Olmschenk09}.  Each ion is entangled with the frequency of a single photon, and the two photons then interfere on a beamsplitter (BS).  Simultaneous detection of photons at photomultiplier tubes (PMTs) after polarization filtering (PBS) heralds ion-ion entanglement, a resource for teleportation.  The $\chi$-matrix  is determined from measurements of six teleported states in three orthogonal photon bases.  Absolute values of the matrix entries are plotted, with rows and columns labeled by the Pauli operators $\{\hat{\sigma}_0,\hat{\sigma}_1,\hat{\sigma}_2,\hat{\sigma}_3\}$. The $(\hat{\sigma}_0,\hat{\sigma}_0)$ entry corresponds to a process fidelity $F = 84(2)\%$. 
\textbf{b} The relationship between $\chi$ and the Bloch/Poincar\'e sphere picture is shown for the transfer of a quantum state from an ion to a cavity photon \cite{Stute13}.  The process maps the pure states on the surface of the sphere to a set of states which are slightly deformed, i.e., reduced in amplitude and rotated, corresponding to a process fidelity $F = 92(2)\%$.
\textbf{c} Another method to evaluate a quantum process is to measure correlations that provide information about coherence.  In Ref. \onlinecite{Hofmann12}, simultaneous detection of two photons projects atoms in two separate traps into a maximally entangled Bell state.  This Bell state measurement (BSM) combines the photons on a fiber beamsplitter (BS), after which they are rotated by half- and quarter-wave plates $\lambda/2,~\lambda/4$ and detected on avalanche photodiodes (APDs) at the horizontal (H) and vertical (V) ports of a polarizing beamsplitter.  Entanglement is verified by reading out the atoms' internal spin states and determining the probability that the spins are correlated or anticorrelated.  Oscillations in the correlation probability as the spin basis for one atom is varied (parameterized by an angle $\beta$) are shown here for the Bell state $|\Psi^+\rangle$, and together with further measurements can be used to estimate a state fidelity  $F = 81(2)\%$ with respect to $|\Psi^+\rangle$.}
 \end{figure}

Mathematically, a quantum process is fully described by the process matrix $\chi$,  which represents the mapping between input and output density matrices, $\rho_{\text{in}} \mapsto \rho_{\text{out}}$: 
\begin{align*}
\rho_\mathrm{out} = \sum_{i,j} \limits \chi_{ij} A_i \rho_\mathrm{in} A_j^{\dagger}, 
\end{align*}
where the operators $A_i$ comprise a basis for operators on the Hilbert space.  (For systems comprised of qubits, it is convenient to define $A_i$ as tensor products of the Pauli operators.)
Since the ideal quantum-information transfer process leaves $\rho_\text{in}$ unchanged, the process fidelity is defined as $F = \text{tr}\sqrt{\rho_{\text{in}}^{1/2} \rho_{\text{out}} \rho_{\text{in}}^{1/2}} = \chi_{00}$.  A value $F = 1$ corresponds to perfect transfer, while $50\%$ represents the classical limit.  For $N$ qubits, $\chi$ can be characterized by sampling the Hilbert space with $4^N$ inputs and measuring each output in $3^N$ bases \cite{Nielsen2000}, as illustrated with a single qubit in Fig. \ref{fig4}.

Process tomography provides a complete picture but is not required in order to establish that the fidelity is nonclassical;
an appropriate set of correlation measurements is sufficient.  
In Ref. \onlinecite{Hofmann12}, for example, the entanglement of two rubidium atoms over 20 m was analyzed based on correlation measurements of fluorescence from the atoms (Fig. \ref{fig4}).  
The measurement basis of each atom was determined by the linear polarization of lasers that transferred population between Zeeman states.  
By rotating the basis of one atom with respect to the other, the authors observed oscillations in the correlation probability sufficient to violate Bell's inequality.

\section{Long-distance transfer}

Fidelity measures how faithfully a quantum state can be transported; another important question to ask is how far that state can be sent.
In both deterministic and heralded schemes, we have seen that photons act as carriers of quantum information.  
The deterministic case is straightforward: a photon physically transports information from one site to another.  
The heralded case is more abstract: teleportation is only possible because two photon paths from remote atoms converged at a common location.  
In either case, if we are interested in long-distance transfer, we must take into account the losses inherent in optical channels.
The classical approach in a fiber-optic network is to intersperse the channel with repeater stations at which the signal is amplified, but in the quantum-mechanical world, this amplification is forbidden by the no-cloning theorem \cite{Wootters1982}.

Instead, the solution is a quantum repeater, by means of which entanglement over relatively short distances is first purified, then extended to increasingly longer distances via entanglement swapping \cite{Briegel98}.  
In order to build up entanglement between successive repeater nodes, quantum memories are essential: they allow quantum states to be stored at one node until an entangled pair is generated at the next node.
A quantum repeater scheme based on atomic ensembles was first proposed in 2001 \cite{Duan01}, and since then, significant progress has been made toward its experimental realization \cite{Sangouard11}, including the demonstration of elementary repeater segments \cite{Chou07, Yuan08}.  In fact, repeater architectures based on all of the quantum memories discussed above, including ions \cite{Monroe13} and NV centers \cite{Childress06}, are being pursued in several groups worldwide.  It is an indication of the complexity of these architectures that entanglement swapping between quantum memories has not yet been achieved.

A central challenge faced in all of these experiments is that the rate of entanglement generation between repeater nodes must be faster than the rate at which this entanglement decoheres.  Otherwise, if entanglement is lost due to interactions with the environment before it can be harnessed, it will not be possible to extend the range of quantum networks via entanglement swapping.   This problem can be addressed in two ways: by speeding up the rate of entanglement generation, and by increasing the storage times of quantum memories.  

The particular approach to speeding up entanglement generation will be specific to each implementation, but typically, photon collection efficiency is the bottleneck in these schemes.  
Schemes based on spontaneous emission can be improved by integrating specialized objectives. 
For ions and neutral atoms, such optics may include in-vacuum lens systems with a high numerical aperture, 
parabolic or spherical mirrors \cite{Maiwald09,Shu10}, or Fresnel optics \cite{Streed11}.
In order to scale up the number of atoms or ions at a given node, it will be important to integrate scalable collection optics, such as microfabricated lenses \cite{Brady11,Merrill11} and optical fibers \cite{VanDevender10}.
For single-photon emission from NV centers, solid-immersion lenses fabricated around the site have been shown to enhance collection efficiency by an order of magnitude \cite{Hadden10,Robledo11}. 
Finally, enclosing the emitters in cavities enables efficient collection of photons into a well-defined optical mode \cite{Ritter12,Casabone13}.  

The second route toward scalable quantum repeaters is to enhance quantum memory storage times, which are often limited by environmental fluctuations, such as drifting magnetic fields.  These parameters can be actively stabilized, but a more robust solution is to use memories that are well isolated from the environment.  For experiments using atomic ensembles as a storage medium, thermal diffusion and collisions are key decoherence channels, and so one solution would be to shift to quantum-degenerate gases, in which both mechanisms are suppressed \cite{Zhang09,Schnorrberger09,Lettner11}.  In ion traps, it is possible to confine multiple atomic species simultaneously, and there are advantages to distributing tasks between two species, which are coupled by shared motional modes \cite{Schmidt05}.  For example, by using one species as a memory and a second species for a quantum interface, the memory is then shielded from laser interactions \cite{Monroe13}. 
This same separation between memory and communication tasks is promising for NV centers, where information can be transferred from the electronic spin state to and from a well-isolated nuclear spin nearby \cite{Maurer12}.
Quite generally, in identifying the particular states of a quantum system in which information is stored, it may be useful to work in decoherence-free subspaces, special states that are decoupled from noise channels \cite{Lidar03}.



\section{On the horizon}

\begin{figure}
\includegraphics[width=0.33\textwidth]{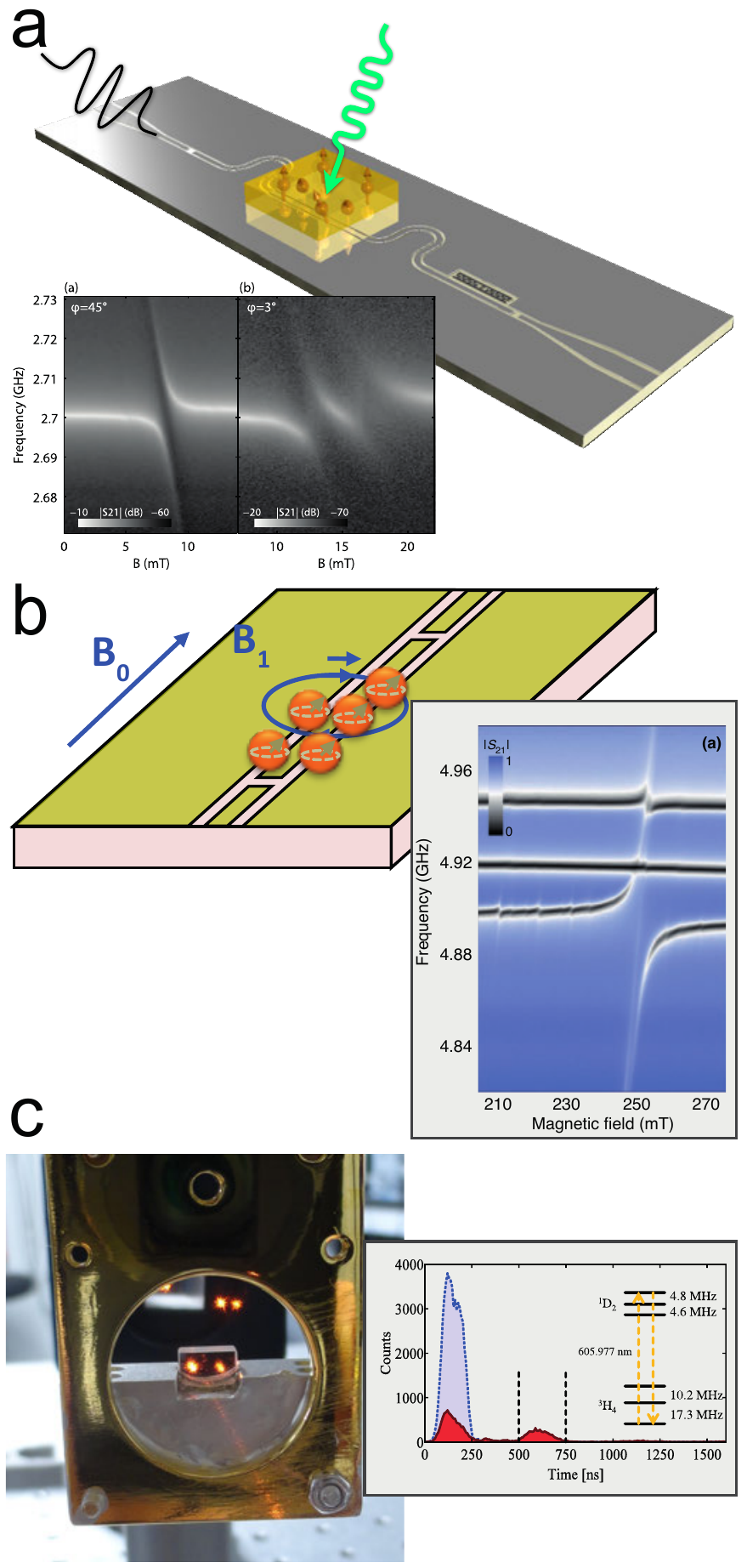}
\caption{\label{fig5}  Several platforms offer promising new approaches to quantum information transfer. \textbf{a} Circuit-QED systems have emerged as a versatile platform for quantum information processing \cite{Devoret13}, but microwave photons are not compatible with long-distance optical channels.  One solution is to construct a microwave-to-optical interface by coupling these superconducting systems to NV centers in diamond, which have transitions in the optical regime.  A microwave stripline resonator is driven in the plane of a chip with embedded NV centers, with optical access from above.  Strong coupling is demonstrated by avoided crossings (vacuum Rabi splittings) as a magnetic field tunes the system into resonance.  Image courtesy of J. Majer, TU Wien/Atominstitut and adapted from Ref.~\onlinecite{Amsuess11}.  \textbf{b} Microwave resonators can also be coupled to crystals doped with rare-earth ions, which have transition frequencies in the telecom band and excellent coherence properties.  Here, Er$^{3+}$ ions are illustrated as a collection of spins above a lumped element resonator, and an avoided crossing is measured.  Image courtesy of P. Bushev, Saarland University and adapted from Ref. \onlinecite{Probst13}.  \textbf{c} Rare-earth ions are also excellent candidates for quantum memories.  Polarization-based photonic qubits are mapped in and out of a Pr$^{3+}$:Y$_2$SiO$_5$ crystal cooled to 2.8K, where the plot shows the storage and retrieval of a vertically polarized qubit.  Image courtesy of H. de Riedmatten, ICFO-Institut de Ci\`{e}ncies Fot\`{o}niques and adapted from Ref. \onlinecite{Guendogan12}. 
}
\end{figure}

Quantum network experiments are still in the proof-of-principle stage, and further development of existing interfaces based on atoms, ions, and solid-state devices will bring substantial gains in both speed and fidelity.
In addition, new hybrid systems are emerging that may allow us to combine the advantages of different platforms (Fig. \ref{fig5}).
For example, quantum interfaces typically interact with photons in the visible or even ultraviolet range, but telecom frequencies in the infrared are better suited to long-distance communication because optical-fiber losses in these bands are minimized.  
An important line of research is thus the efficient conversion of photons to and from telecom frequencies in order to extend the range of quantum network channels \cite{Radnaev10,Degreve12}.

Another hybrid system comes from the field of circuit quantum electrodynamics (circuit QED), based on Josephson-junction artificial atoms coupled to superconducting microwave resonators.  In the past decade, circuit-QED devices have proven to be strong candidate systems for quantum computing \cite{Devoret13}.  Recent Hong-Ou-Mandel experiments have shown that these devices can produce indistinguishable photons as required for remote entanglement \cite{Lang13}, but since microwave photons are not well suited for long-distance communication, an optical-to-microwave interface is needed.  Potential routes toward such an interface can be found in recent work that couples a superconducting qubit to optical spin ensembles, such as NV centers \cite{Kubo10,Schuster10,Zhu11,Amsuess11} or crystals doped with rare-earth ions \cite{Probst13}.  

We have already outlined the contrast between quantum nodes, which are essentially local quantum computers,
and quantum memories, which focus on on information storage and play an essential role in quantum computers.  
Rare-earth ions show particular promise as quantum memories, as they offer long coherence times \cite{Longdell05} and efficient storage \cite{Hedges10} within a solid-state platform, interactions at telecom wavelengths, and multiplexed storage protocols that could significantly speed up long-distance entanglement rates \cite{Simon07}.  Storage of photonic polarization quantum bits has recently been demonstrated with both high efficiency and fidelity, paving the way for interactions with polarization-based quantum-information processors  \cite{Clausen12,Guendogan12,Zhou12}.

This review has highlighted how optical cavities enable light--matter information transfer.  
Another role for optical quantum memories in cavities has recently emerged: such systems can act as transistors, that is, as single-photon gates for light fields \cite{Chen13} that may be integrated in future quantum networks.

Quantum information science is still a young field; as we have seen, it is just within the past decade that the first realizations of photon-based transfer have been achieved.  As progress continues, the crucial challenge will be to find systems where the efficiency and fidelity of information transfer are sufficient to implement error correction, the basis for a robust and scalable network.

\makeatletter
 \def\@biblabel#1{#1}
\makeatother

\bibliography{cqed_bibsonomy}

\begin{thebibliography}{100}
\expandafter\ifx\csname url\endcsname\relax
  \def\url#1{\texttt{#1}}\fi
\expandafter\ifx\csname urlprefix\endcsname\relax\def\urlprefix{URL }\fi
\providecommand{\bibinfo}[2]{#2}
\providecommand{\eprint}[2][]{\url{#2}}

\bibitem{Gisin07}
\bibinfo{author}{Gisin, N.} \& \bibinfo{author}{Thew, R.}
\newblock \bibinfo{title}{Quantum communication}.
\newblock \emph{\bibinfo{journal}{Nature Photon.}}
  \textbf{\bibinfo{volume}{1}}, \bibinfo{pages}{165--171}
  (\bibinfo{year}{2007}).

\bibitem{Kimble08a}
\bibinfo{author}{Kimble, H.J.}
\newblock \bibinfo{title}{{The quantum internet}}.
\newblock \emph{\bibinfo{journal}{Nature}} \textbf{\bibinfo{volume}{453}},
  \bibinfo{pages}{1023--1030} (\bibinfo{year}{2008}).

\bibitem{Yin12}
\bibinfo{author}{Yin, J.} \emph{et~al.}
\newblock \bibinfo{title}{Quantum teleportation and entanglement distribution
  over 100-kilometre free-space channels}.
\newblock \emph{\bibinfo{journal}{Nature}} \textbf{\bibinfo{volume}{488}},
  \bibinfo{pages}{185--188} (\bibinfo{year}{2012}).

\bibitem{Ma12}
\bibinfo{author}{Ma, X.S.} \emph{et~al.}
\newblock \bibinfo{title}{Quantum teleportation over 143 kilometres using
  active feed-forward}.
\newblock \emph{\bibinfo{journal}{Nature}} \textbf{\bibinfo{volume}{489}},
  \bibinfo{pages}{269--273} (\bibinfo{year}{2012}).

\bibitem{Gisin02}
\bibinfo{author}{Gisin, N.}, \bibinfo{author}{Ribordy, G.},
  \bibinfo{author}{Tittel, W.} \& \bibinfo{author}{Zbinden, H.}
\newblock \bibinfo{title}{Quantum cryptography}.
\newblock \emph{\bibinfo{journal}{Rev. Mod. Phys.}}
  \textbf{\bibinfo{volume}{74}}, \bibinfo{pages}{145--195}
  (\bibinfo{year}{2002}).

\bibitem{Shor94}
\bibinfo{author}{Shor, P.W.}
\newblock \bibinfo{title}{Algorithms for quantum computation: discrete
  logarithms and factoring}.
\newblock In \emph{\bibinfo{booktitle}{Proc. Annu. Symp. Found. Comput. Sci.}},
  \bibinfo{pages}{124--134} (\bibinfo{year}{1994}).

\bibitem{CiracZoller95}
\bibinfo{author}{Cirac, J.I.} \& \bibinfo{author}{Zoller, P.}
\newblock \bibinfo{title}{Quantum computations with cold trapped ions}.
\newblock \emph{\bibinfo{journal}{Phys. Rev. Lett.}}
  \textbf{\bibinfo{volume}{74}}, \bibinfo{pages}{4091--4094}
  (\bibinfo{year}{1995}).

\bibitem{Cirac97}
\bibinfo{author}{Cirac, J.I.}, \bibinfo{author}{Zoller, P.},
  \bibinfo{author}{Kimble, H.J.} \& \bibinfo{author}{Mabuchi, H.}
\newblock \bibinfo{title}{Quantum state transfer and entanglement distribution
  among distant nodes in a quantum network}.
\newblock \emph{\bibinfo{journal}{Phys. Rev. Lett.}}
  \textbf{\bibinfo{volume}{78}}, \bibinfo{pages}{3221--3224}
  (\bibinfo{year}{1997}).

\bibitem{Haroche06}
\bibinfo{author}{Haroche, S.} \& \bibinfo{author}{Raimond, J.}
\newblock \emph{\bibinfo{title}{Exploring the quantum: atoms, cavities and
  photons}} (\bibinfo{publisher}{Oxford University Press},
  \bibinfo{address}{Oxford, UK}, \bibinfo{year}{2006}).

\bibitem{Brendel99}
\bibinfo{author}{Brendel, J.}, \bibinfo{author}{Gisin, N.},
  \bibinfo{author}{Tittel, W.} \& \bibinfo{author}{Zbinden, H.}
\newblock \bibinfo{title}{Pulsed energy-time entangled twin-photon source for
  quantum communication}.
\newblock \emph{\bibinfo{journal}{Phys. Rev. Lett.}}
  \textbf{\bibinfo{volume}{82}}, \bibinfo{pages}{2594--2597}
  (\bibinfo{year}{1999}).

\bibitem{vanEnk97}
\bibinfo{author}{van Enk, S.J.}, \bibinfo{author}{Cirac, J.I.} \&
  \bibinfo{author}{Zoller, P.}
\newblock \bibinfo{title}{Ideal quantum communication over noisy channels: A
  quantum optical implementation}.
\newblock \emph{\bibinfo{journal}{Phys. Rev. Lett.}}
  \textbf{\bibinfo{volume}{78}}, \bibinfo{pages}{4293--4296}
  (\bibinfo{year}{1997}).

\bibitem{vanEnk97a}
\bibinfo{author}{van Enk, S.J.}, \bibinfo{author}{Cirac, J.I.} \&
  \bibinfo{author}{Zoller, P.}
\newblock \bibinfo{title}{Purifying two-bit quantum gates and joint
  measurements in cavity {QED}}.
\newblock \emph{\bibinfo{journal}{Phys. Rev. Lett.}}
  \textbf{\bibinfo{volume}{79}}, \bibinfo{pages}{5178--5181}
  (\bibinfo{year}{1997}).

\bibitem{Duan10}
\bibinfo{author}{Duan, L.M.} \& \bibinfo{author}{Monroe, C.}
\newblock \bibinfo{title}{Colloquium: Quantum networks with trapped ions}.
\newblock \emph{\bibinfo{journal}{Rev. Mod. Phys.}}
  \textbf{\bibinfo{volume}{82}}, \bibinfo{pages}{1209--1224}
  (\bibinfo{year}{2010}).

\bibitem{Monroe13}
\bibinfo{author}{Monroe, C.} \& \bibinfo{author}{Kim, J.}
\newblock \bibinfo{title}{Scaling the ion trap quantum processor}.
\newblock \emph{\bibinfo{journal}{Science}} \textbf{\bibinfo{volume}{339}},
  \bibinfo{pages}{1164--1169} (\bibinfo{year}{2013}).

\bibitem{Kuhn99}
\bibinfo{author}{{Kuhn}, A.}, \bibinfo{author}{{Hennrich}, M.},
  \bibinfo{author}{{Bondo}, T.} \& \bibinfo{author}{{Rempe}, G.}
\newblock \bibinfo{title}{{Controlled generation of single photons from a
  strongly coupled atom-cavity system}}.
\newblock \emph{\bibinfo{journal}{Appl. Phys. B}}
  \textbf{\bibinfo{volume}{69}}, \bibinfo{pages}{373--377}
  (\bibinfo{year}{1999}).

\bibitem{Keller04}
\bibinfo{author}{Keller, M.}, \bibinfo{author}{Lange, B.},
  \bibinfo{author}{Hayasaka, K.}, \bibinfo{author}{Lange, W.} \&
  \bibinfo{author}{Walther, H.}
\newblock \bibinfo{title}{Continuous generation of single photons with
  controlled waveform in an ion-trap cavity system}.
\newblock \emph{\bibinfo{journal}{Nature}} \textbf{\bibinfo{volume}{431}},
  \bibinfo{pages}{1075--1078} (\bibinfo{year}{2004}).

\bibitem{Cabrillo99}
\bibinfo{author}{Cabrillo, C.}, \bibinfo{author}{Cirac, J.I.},
  \bibinfo{author}{Garc\'{i}a-Fern\'{a}ndez, P.} \& \bibinfo{author}{Zoller,
  P.}
\newblock \bibinfo{title}{Creation of entangled states of distant atoms by
  interference}.
\newblock \emph{\bibinfo{journal}{Phys. Rev. A}} \textbf{\bibinfo{volume}{59}},
  \bibinfo{pages}{1025--1033} (\bibinfo{year}{1999}).

\bibitem{Feng03}
\bibinfo{author}{Feng, X.L.}, \bibinfo{author}{Zhang, Z.M.},
  \bibinfo{author}{Li, X.D.}, \bibinfo{author}{Gong, S.Q.} \&
  \bibinfo{author}{Xu, Z.Z.}
\newblock \bibinfo{title}{Entangling distant atoms by interference of polarized
  photons}.
\newblock \emph{\bibinfo{journal}{Phys. Rev. Lett.}}
  \textbf{\bibinfo{volume}{90}}, \bibinfo{pages}{217902}
  (\bibinfo{year}{2003}).

\bibitem{Duan03}
\bibinfo{author}{Duan, L.M.} \& \bibinfo{author}{Kimble, H.J.}
\newblock \bibinfo{title}{Efficient engineering of multiatom entanglement
  through single-photon detections}.
\newblock \emph{\bibinfo{journal}{Phys. Rev. Lett.}}
  \textbf{\bibinfo{volume}{90}}, \bibinfo{pages}{253601}
  (\bibinfo{year}{2003}).

\bibitem{Simon03}
\bibinfo{author}{Simon, C.} \& \bibinfo{author}{Irvine, W.T.M.}
\newblock \bibinfo{title}{Robust long-distance entanglement and a loophole-free
  {B}ell test with ions and photons}.
\newblock \emph{\bibinfo{journal}{Phys. Rev. Lett.}}
  \textbf{\bibinfo{volume}{91}}, \bibinfo{pages}{110405}
  (\bibinfo{year}{2003}).

\bibitem{Bennett93}
\bibinfo{author}{Bennett, C.H.} \emph{et~al.}
\newblock \bibinfo{title}{Teleporting an unknown quantum state via dual
  classical and {E}instein-{P}odolsky-{R}osen channels}.
\newblock \emph{\bibinfo{journal}{Phys. Rev. Lett.}}
  \textbf{\bibinfo{volume}{70}}, \bibinfo{pages}{1895--1899}
  (\bibinfo{year}{1993}).

\bibitem{Ye99}
\bibinfo{author}{Ye, J.}, \bibinfo{author}{Vernooy, D.W.} \&
  \bibinfo{author}{Kimble, H.J.}
\newblock \bibinfo{title}{Trapping of single atoms in cavity {QED}}.
\newblock \emph{\bibinfo{journal}{Phys. Rev. Lett.}}
  \textbf{\bibinfo{volume}{83}}, \bibinfo{pages}{4987--4990}
  (\bibinfo{year}{1999}).

\bibitem{Puppe07}
\bibinfo{author}{Puppe, T.} \emph{et~al.}
\newblock \bibinfo{title}{Trapping and observing single atoms in a blue-detuned
  intracavity dipole trap}.
\newblock \emph{\bibinfo{journal}{Phys. Rev. Lett.}}
  \textbf{\bibinfo{volume}{99}}, \bibinfo{pages}{013002}
  (\bibinfo{year}{2007}).

\bibitem{Sauer04a}
\bibinfo{author}{Sauer, J.A.}, \bibinfo{author}{Fortier, K.M.},
  \bibinfo{author}{Chang, M.S.}, \bibinfo{author}{Hamley, C.D.} \&
  \bibinfo{author}{Chapman, M.S.}
\newblock \bibinfo{title}{Cavity {QED} with optically transported atoms}.
\newblock \emph{\bibinfo{journal}{Phys. Rev. A}} \textbf{\bibinfo{volume}{69}},
  \bibinfo{pages}{051804} (\bibinfo{year}{2004}).

\bibitem{Nussmann05}
\bibinfo{author}{Nu{\ss}mann, S.} \emph{et~al.}
\newblock \bibinfo{title}{Submicron positioning of single atoms in a
  microcavity}.
\newblock \emph{\bibinfo{journal}{Phys. Rev. Lett.}}
  \textbf{\bibinfo{volume}{95}}, \bibinfo{pages}{173602}
  (\bibinfo{year}{2005}).

\bibitem{Khudaverdyan09}
\bibinfo{author}{Khudaverdyan, M.} \emph{et~al.}
\newblock \bibinfo{title}{Quantum jumps and spin dynamics of interacting atoms
  in a strongly coupled atom-cavity system}.
\newblock \emph{\bibinfo{journal}{Phys. Rev. Lett.}}
  \textbf{\bibinfo{volume}{103}}, \bibinfo{pages}{123006}
  (\bibinfo{year}{2009}).

\bibitem{Ritter12}
\bibinfo{author}{Ritter, S.} \emph{et~al.}
\newblock \bibinfo{title}{An elementary quantum network of single atoms in
  optical cavities}.
\newblock \emph{\bibinfo{journal}{Nature}} \textbf{\bibinfo{volume}{484}},
  \bibinfo{pages}{195--200} (\bibinfo{year}{2012}).

\bibitem{Schrader01}
\bibinfo{author}{Schrader, D.} \emph{et~al.}
\newblock \bibinfo{title}{An optical conveyor belt for single neutral atoms}.
\newblock \emph{\bibinfo{journal}{Appl. Phys. B}}
  \textbf{\bibinfo{volume}{73}}, \bibinfo{pages}{819--824}
  (\bibinfo{year}{2001}).

\bibitem{Miller05}
\bibinfo{author}{Miller, R.} \emph{et~al.}
\newblock \bibinfo{title}{Trapped atoms in cavity {QED}: coupling quantized
  light and matter}.
\newblock \emph{\bibinfo{journal}{J. Phys. B}} \textbf{\bibinfo{volume}{38}},
  \bibinfo{pages}{S551--S565} (\bibinfo{year}{2005}).

\bibitem{Paul90}
\bibinfo{author}{Paul, W.}
\newblock \bibinfo{title}{Electromagnetic traps for charged and neutral
  particles}.
\newblock \emph{\bibinfo{journal}{Rev. Mod. Phys.}}
  \textbf{\bibinfo{volume}{62}}, \bibinfo{pages}{531--540}
  (\bibinfo{year}{1990}).

\bibitem{Steiner13}
\bibinfo{author}{Steiner, M.}, \bibinfo{author}{Meyer, H.M.},
  \bibinfo{author}{Deutsch, C.}, \bibinfo{author}{Reichel, J.} \&
  \bibinfo{author}{K\"ohl, M.}
\newblock \bibinfo{title}{Single ion coupled to an optical fiber cavity}.
\newblock \emph{\bibinfo{journal}{Phys. Rev. Lett.}}
  \textbf{\bibinfo{volume}{110}}, \bibinfo{pages}{043003}
  (\bibinfo{year}{2013}).

\bibitem{Guthoehrlein01}
\bibinfo{author}{Guth\"{o}hrlein, G.R.}, \bibinfo{author}{Keller, M.},
  \bibinfo{author}{Hayasaka, K.}, \bibinfo{author}{Lange, W.} \&
  \bibinfo{author}{Walther, H.}
\newblock \bibinfo{title}{A single ion as a nanoscopic probe of an optical
  field}.
\newblock \emph{\bibinfo{journal}{Nature}} \textbf{\bibinfo{volume}{414}},
  \bibinfo{pages}{49--51} (\bibinfo{year}{2001}).

\bibitem{Russo09}
\bibinfo{author}{Russo, C.} \emph{et~al.}
\newblock \bibinfo{title}{Raman spectroscopy of a single ion coupled to a
  high-finesse cavity}.
\newblock \emph{\bibinfo{journal}{Appl. Phys. B}}
  \textbf{\bibinfo{volume}{95}}, \bibinfo{pages}{205--212}
  (\bibinfo{year}{2009}).

\bibitem{Boozer06}
\bibinfo{author}{Boozer, A.D.}, \bibinfo{author}{Boca, A.},
  \bibinfo{author}{Miller, R.}, \bibinfo{author}{Northup, T.E.} \&
  \bibinfo{author}{Kimble, H.J.}
\newblock \bibinfo{title}{Cooling to the ground state of axial motion for one
  atom strongly coupled to an optical cavity}.
\newblock \emph{\bibinfo{journal}{Phys. Rev. Lett.}}
  \textbf{\bibinfo{volume}{97}}, \bibinfo{pages}{083602}
  (\bibinfo{year}{2006}).

\bibitem{Stute12a}
\bibinfo{author}{Stute, A.} \emph{et~al.}
\newblock \bibinfo{title}{Toward an ion-photon quantum interface in an optical
  cavity}.
\newblock \emph{\bibinfo{journal}{Appl. Phys. B}}
  \textbf{\bibinfo{volume}{107}}, \bibinfo{pages}{1145--1157}
  (\bibinfo{year}{2012}).

\bibitem{Reiserer13}
\bibinfo{author}{Reiserer, A.}, \bibinfo{author}{N\"olleke, C.},
  \bibinfo{author}{Ritter, S.} \& \bibinfo{author}{Rempe, G.}
\newblock \bibinfo{title}{Ground-state cooling of a single atom at the center
  of an optical cavity}.
\newblock \emph{\bibinfo{journal}{Phys. Rev. Lett.}}
  \textbf{\bibinfo{volume}{110}}, \bibinfo{pages}{223003}
  (\bibinfo{year}{2013}).

\bibitem{Leibfried03}
\bibinfo{author}{Leibfried, D.}, \bibinfo{author}{Blatt, R.},
  \bibinfo{author}{Monroe, C.} \& \bibinfo{author}{Wineland, D.}
\newblock \bibinfo{title}{Quantum dynamics of single trapped ions}.
\newblock \emph{\bibinfo{journal}{Rev. Mod. Phys.}}
  \textbf{\bibinfo{volume}{75}}, \bibinfo{pages}{281--324}
  (\bibinfo{year}{2003}).

\bibitem{Horak97}
\bibinfo{author}{Horak, P.}, \bibinfo{author}{Hechenblaikner, G.},
  \bibinfo{author}{Gheri, K.M.}, \bibinfo{author}{Stecher, H.} \&
  \bibinfo{author}{Ritsch, H.}
\newblock \bibinfo{title}{Cavity-induced atom cooling in the strong coupling
  regime}.
\newblock \emph{\bibinfo{journal}{Phys. Rev. Lett.}}
  \textbf{\bibinfo{volume}{79}}, \bibinfo{pages}{4974--4977}
  (\bibinfo{year}{1997}).

\bibitem{Chan03}
\bibinfo{author}{Chan, H.W.}, \bibinfo{author}{Black, A.T.} \&
  \bibinfo{author}{Vuleti\'{c}, V.}
\newblock \bibinfo{title}{Observation of collective-emission-induced cooling of
  atoms in an optical cavity}.
\newblock \emph{\bibinfo{journal}{Phys. Rev. Lett.}}
  \textbf{\bibinfo{volume}{90}}, \bibinfo{pages}{063003}
  (\bibinfo{year}{2003}).

\bibitem{Maunz04}
\bibinfo{author}{Maunz, P.} \emph{et~al.}
\newblock \bibinfo{title}{Cavity cooling of a single atom}.
\newblock \emph{\bibinfo{journal}{Nature}} \textbf{\bibinfo{volume}{428}},
  \bibinfo{pages}{50--52} (\bibinfo{year}{2004}).

\bibitem{Fortier07}
\bibinfo{author}{Fortier, K.M.}, \bibinfo{author}{Kim, S.Y.},
  \bibinfo{author}{Gibbons, M.J.}, \bibinfo{author}{Ahmadi, P.} \&
  \bibinfo{author}{Chapman, M.S.}
\newblock \bibinfo{title}{Deterministic loading of individual atoms to a
  high-finesse optical cavity}.
\newblock \emph{\bibinfo{journal}{Phys. Rev. Lett.}}
  \textbf{\bibinfo{volume}{98}}, \bibinfo{pages}{233601}
  (\bibinfo{year}{2007}).

\bibitem{Leibrandt09}
\bibinfo{author}{Leibrandt, D.R.}, \bibinfo{author}{Labaziewicz, J.},
  \bibinfo{author}{Vuleti\'{c}, V.} \& \bibinfo{author}{Chuang, I.L.}
\newblock \bibinfo{title}{Cavity sideband cooling of a single trapped ion}.
\newblock \emph{\bibinfo{journal}{Phys. Rev. Lett.}}
  \textbf{\bibinfo{volume}{103}}, \bibinfo{pages}{103001}
  (\bibinfo{year}{2009}).

\bibitem{Kubanek09}
\bibinfo{author}{Kubanek, A.} \emph{et~al.}
\newblock \bibinfo{title}{{Photon-by-photon feedback control of a single-atom
  trajectory}}.
\newblock \emph{\bibinfo{journal}{Nature}} \textbf{\bibinfo{volume}{462}},
  \bibinfo{pages}{898--901} (\bibinfo{year}{2009}).

\bibitem{Koch10}
\bibinfo{author}{Koch, M.} \emph{et~al.}
\newblock \bibinfo{title}{Feedback cooling of a single neutral atom}.
\newblock \emph{\bibinfo{journal}{Phys. Rev. Lett.}}
  \textbf{\bibinfo{volume}{105}}, \bibinfo{pages}{173003}
  (\bibinfo{year}{2010}).

\bibitem{Boozer07a}
\bibinfo{author}{Boozer, A.D.}, \bibinfo{author}{Boca, A.},
  \bibinfo{author}{Miller, R.}, \bibinfo{author}{Northup, T.E.} \&
  \bibinfo{author}{Kimble, H.J.}
\newblock \bibinfo{title}{Reversible state transfer between light and a single
  trapped atom}.
\newblock \emph{\bibinfo{journal}{Phys. Rev. Lett.}}
  \textbf{\bibinfo{volume}{98}}, \bibinfo{pages}{193601}
  (\bibinfo{year}{2007}).

\bibitem{Wilk07b}
\bibinfo{author}{Wilk, T.}, \bibinfo{author}{Webster, S.C.},
  \bibinfo{author}{Kuhn, A.} \& \bibinfo{author}{Rempe, G.}
\newblock \bibinfo{title}{Single-atom single-photon quantum interface}.
\newblock \emph{\bibinfo{journal}{Science}} \textbf{\bibinfo{volume}{317}},
  \bibinfo{pages}{488--490} (\bibinfo{year}{2007}).

\bibitem{Specht11}
\bibinfo{author}{Specht, H.P.} \emph{et~al.}
\newblock \bibinfo{title}{A single-atom quantum memory}.
\newblock \emph{\bibinfo{journal}{Nature}} \textbf{\bibinfo{volume}{473}},
  \bibinfo{pages}{190--193} (\bibinfo{year}{2011}).

\bibitem{Lettner11}
\bibinfo{author}{Lettner, M.} \emph{et~al.}
\newblock \bibinfo{title}{Remote entanglement between a single atom and a
  {B}ose-{E}instein condensate}.
\newblock \emph{\bibinfo{journal}{Phys. Rev. Lett.}}
  \textbf{\bibinfo{volume}{106}}, \bibinfo{pages}{210503}
  (\bibinfo{year}{2011}).

\bibitem{Stute13}
\bibinfo{author}{Stute, A.} \emph{et~al.}
\newblock \bibinfo{title}{Quantum-state transfer from an ion to a photon}.
\newblock \emph{\bibinfo{journal}{Nature Photon.}}
  \textbf{\bibinfo{volume}{7}}, \bibinfo{pages}{219--222}
  (\bibinfo{year}{2013}).

\bibitem{Haeffner08}
\bibinfo{author}{H{\"a}ffner, H.}, \bibinfo{author}{Roos, C.} \&
  \bibinfo{author}{Blatt, R.}
\newblock \bibinfo{title}{{Quantum computing with trapped ions}}.
\newblock \emph{\bibinfo{journal}{Phys. Rep.}} \textbf{\bibinfo{volume}{469}},
  \bibinfo{pages}{155--203} (\bibinfo{year}{2008}).

\bibitem{Blinov04}
\bibinfo{author}{Blinov, B.B.}, \bibinfo{author}{Moehring, D.L.},
  \bibinfo{author}{Duan, L.M.} \& \bibinfo{author}{Monroe, C.}
\newblock \bibinfo{title}{Observation of entanglement between a single trapped
  atom and a single photon}.
\newblock \emph{\bibinfo{journal}{Nature}} \textbf{\bibinfo{volume}{428}},
  \bibinfo{pages}{153--157} (\bibinfo{year}{2004}).

\bibitem{Olmschenk09}
\bibinfo{author}{Olmschenk, S.} \emph{et~al.}
\newblock \bibinfo{title}{Quantum teleportation between distant matter qubits}.
\newblock \emph{\bibinfo{journal}{Science}} \textbf{\bibinfo{volume}{323}},
  \bibinfo{pages}{486--489} (\bibinfo{year}{2009}).

\bibitem{Volz06}
\bibinfo{author}{Volz, J.} \emph{et~al.}
\newblock \bibinfo{title}{Observation of entanglement of a single photon with a
  trapped atom}.
\newblock \emph{\bibinfo{journal}{Phys. Rev. Lett.}}
  \textbf{\bibinfo{volume}{96}}, \bibinfo{pages}{030404}
  (\bibinfo{year}{2006}).

\bibitem{Matsukevich05}
\bibinfo{author}{Matsukevich, D.N.} \emph{et~al.}
\newblock \bibinfo{title}{Entanglement of a photon and a collective atomic
  excitation}.
\newblock \emph{\bibinfo{journal}{Phys. Rev. Lett.}}
  \textbf{\bibinfo{volume}{95}}, \bibinfo{pages}{040405}
  (\bibinfo{year}{2005}).

\bibitem{Sherson06}
\bibinfo{author}{Sherson, J.} \emph{et~al.}
\newblock \bibinfo{title}{Quantum teleportation between light and matter}.
\newblock \emph{\bibinfo{journal}{Nature}} \textbf{\bibinfo{volume}{443}},
  \bibinfo{pages}{557--560} (\bibinfo{year}{2006}).

\bibitem{Togan10}
\bibinfo{author}{Togan, E.} \emph{et~al.}
\newblock \bibinfo{title}{Quantum entanglement between an optical photon and a
  solid-state spin qubit}.
\newblock \emph{\bibinfo{journal}{Nature}} \textbf{\bibinfo{volume}{466}},
  \bibinfo{pages}{730--734} (\bibinfo{year}{2010}).

\bibitem{Lee11}
\bibinfo{author}{Lee, K.C.} \emph{et~al.}
\newblock \bibinfo{title}{Entangling macroscopic diamonds at room temperature}.
\newblock \emph{\bibinfo{journal}{Science}} \textbf{\bibinfo{volume}{334}},
  \bibinfo{pages}{1253--1256} (\bibinfo{year}{2011}).

\bibitem{Gao12}
\bibinfo{author}{Gao, W.B.}, \bibinfo{author}{Fallahi, P.},
  \bibinfo{author}{Togan, E.}, \bibinfo{author}{Miguel-Sanchez, J.} \&
  \bibinfo{author}{Imamoglu, A.}
\newblock \bibinfo{title}{Observation of entanglement between a quantum dot
  spin and a single photon}.
\newblock \emph{\bibinfo{journal}{Nature}} \textbf{\bibinfo{volume}{491}},
  \bibinfo{pages}{426--430} (\bibinfo{year}{2012}).

\bibitem{Degreve12}
\bibinfo{author}{De~Greve, K.} \emph{et~al.}
\newblock \bibinfo{title}{Quantum-dot spin-photon entanglement via frequency
  downconversion to telecom wavelength}.
\newblock \emph{\bibinfo{journal}{Nature}} \textbf{\bibinfo{volume}{491}},
  \bibinfo{pages}{421--425} (\bibinfo{year}{2012}).

\bibitem{Schaibley13}
\bibinfo{author}{Schaibley, J.R.} \emph{et~al.}
\newblock \bibinfo{title}{Demonstration of quantum entanglement between a
  single electron spin confined to an {I}n{A}s quantum dot and a photon}.
\newblock \emph{\bibinfo{journal}{Phys. Rev. Lett.}}
  \textbf{\bibinfo{volume}{110}}, \bibinfo{pages}{167401}
  (\bibinfo{year}{2013}).

\bibitem{Stute12}
\bibinfo{author}{Stute, A.} \emph{et~al.}
\newblock \bibinfo{title}{Tunable ion-photon entanglement in an optical
  cavity}.
\newblock \emph{\bibinfo{journal}{Nature}} \textbf{\bibinfo{volume}{485}},
  \bibinfo{pages}{482--485} (\bibinfo{year}{2012}).

\bibitem{Hong87}
\bibinfo{author}{Hong, C.K.}, \bibinfo{author}{Ou, Z.Y.} \&
  \bibinfo{author}{Mandel, L.}
\newblock \bibinfo{title}{Measurement of subpicosecond time intervals between
  two photons by interference}.
\newblock \emph{\bibinfo{journal}{Phys. Rev. Lett.}}
  \textbf{\bibinfo{volume}{59}}, \bibinfo{pages}{2044--2046}
  (\bibinfo{year}{1987}).

\bibitem{Chou05}
\bibinfo{author}{Chou, C.W.} \emph{et~al.}
\newblock \bibinfo{title}{Measurement-induced entanglement for excitation
  stored in remote atomic ensembles}.
\newblock \emph{\bibinfo{journal}{Nature}} \textbf{\bibinfo{volume}{438}},
  \bibinfo{pages}{828--832} (\bibinfo{year}{2005}).

\bibitem{Moehring07}
\bibinfo{author}{Moehring, D.L.} \emph{et~al.}
\newblock \bibinfo{title}{{Entanglement of single-atom quantum bits at a
  distance}}.
\newblock \emph{\bibinfo{journal}{Nature}} \textbf{\bibinfo{volume}{449}},
  \bibinfo{pages}{68--71} (\bibinfo{year}{2007}).

\bibitem{Hofmann12}
\bibinfo{author}{Hofmann, J.} \emph{et~al.}
\newblock \bibinfo{title}{Heralded entanglement between widely separated
  atoms}.
\newblock \emph{\bibinfo{journal}{Science}} \textbf{\bibinfo{volume}{337}},
  \bibinfo{pages}{72--75} (\bibinfo{year}{2012}).

\bibitem{Usmani12}
\bibinfo{author}{Usmani, I.} \emph{et~al.}
\newblock \bibinfo{title}{Heralded quantum entanglement between two crystals}.
\newblock \emph{\bibinfo{journal}{Nature Photon.}}
  \textbf{\bibinfo{volume}{6}}, \bibinfo{pages}{234--237}
  (\bibinfo{year}{2012}).

\bibitem{Bernien13}
\bibinfo{author}{Bernien, H.} \emph{et~al.}
\newblock \bibinfo{title}{Heralded entanglement between solid-state qubits
  separated by three metres}.
\newblock \emph{\bibinfo{journal}{Nature}} \textbf{\bibinfo{volume}{497}},
  \bibinfo{pages}{86--90} (\bibinfo{year}{2013}).

\bibitem{Riebe04}
\bibinfo{author}{Riebe, M.} \emph{et~al.}
\newblock \bibinfo{title}{Deterministic quantum teleportation with atoms}.
\newblock \emph{\bibinfo{journal}{Nature}} \textbf{\bibinfo{volume}{429}},
  \bibinfo{pages}{734--737} (\bibinfo{year}{2004}).

\bibitem{Barrett04}
\bibinfo{author}{Barrett, M.D.} \emph{et~al.}
\newblock \bibinfo{title}{Deterministic quantum teleportation of atomic
  qubits}.
\newblock \emph{\bibinfo{journal}{Nature}} \textbf{\bibinfo{volume}{429}},
  \bibinfo{pages}{737--739} (\bibinfo{year}{2004}).

\bibitem{Maunz09}
\bibinfo{author}{Maunz, P.} \emph{et~al.}
\newblock \bibinfo{title}{Heralded quantum gate between remote quantum
  memories}.
\newblock \emph{\bibinfo{journal}{Phys. Rev. Lett.}}
  \textbf{\bibinfo{volume}{102}}, \bibinfo{pages}{250502}
  (\bibinfo{year}{2009}).

\bibitem{Noelleke13}
\bibinfo{author}{N{\"o}lleke, C.} \emph{et~al.}
\newblock \bibinfo{title}{Efficient teleportation between remote single-atom
  quantum memories}.
\newblock \emph{\bibinfo{journal}{Phys. Rev. Lett.}}
  \textbf{\bibinfo{volume}{110}}, \bibinfo{pages}{140403}
  (\bibinfo{year}{2013}).

\bibitem{Bao12}
\bibinfo{author}{Bao, X.H.} \emph{et~al.}
\newblock \bibinfo{title}{Quantum teleportation between remote atomic-ensemble
  quantum memories}.
\newblock \emph{\bibinfo{journal}{Proc. Natl. Acad. Sci. USA}}
  \textbf{\bibinfo{volume}{109}}, \bibinfo{pages}{20347--20351}
  (\bibinfo{year}{2012}).

\bibitem{Krauter13}
\bibinfo{author}{Krauter, H.} \emph{et~al.}
\newblock \bibinfo{title}{Deterministic quantum teleportation between distant
  atomic objects}.
\newblock \emph{\bibinfo{journal}{Nat. Phys.}} \textbf{\bibinfo{volume}{9}},
  \bibinfo{pages}{400--404} (\bibinfo{year}{2013}).

\bibitem{Nielsen2000}
\bibinfo{author}{Nielsen, M.A.} \& \bibinfo{author}{Chuang, I.L.}
\newblock \emph{\bibinfo{title}{Quantum Computation and Quantum Information}}
  (\bibinfo{publisher}{Cambridge University Press},
  \bibinfo{address}{Cambridge}, \bibinfo{year}{2010}).

\bibitem{Wootters1982}
\bibinfo{author}{Wootters, W.} \& \bibinfo{author}{Zurek, W.}
\newblock \bibinfo{title}{A single quantum cannot be cloned}.
\newblock \emph{\bibinfo{journal}{Nature}} \textbf{\bibinfo{volume}{299}},
  \bibinfo{pages}{802--803} (\bibinfo{year}{1982}).

\bibitem{Briegel98}
\bibinfo{author}{Briegel, H.J.}, \bibinfo{author}{D\"ur, W.},
  \bibinfo{author}{Cirac, J.I.} \& \bibinfo{author}{Zoller, P.}
\newblock \bibinfo{title}{Quantum repeaters: the role of imperfect local
  operations in quantum communication}.
\newblock \emph{\bibinfo{journal}{Phys. Rev. Lett.}}
  \textbf{\bibinfo{volume}{81}}, \bibinfo{pages}{5932--5935}
  (\bibinfo{year}{1998}).

\bibitem{Duan01}
\bibinfo{author}{Duan, L.M.}, \bibinfo{author}{Lukin, M.D.},
  \bibinfo{author}{Cirac, J.I.} \& \bibinfo{author}{Zoller, P.}
\newblock \bibinfo{title}{Long-distance quantum communication with atomic
  ensembles and linear optics}.
\newblock \emph{\bibinfo{journal}{Nature}} \textbf{\bibinfo{volume}{414}},
  \bibinfo{pages}{413--418} (\bibinfo{year}{2001}).

\bibitem{Sangouard11}
\bibinfo{author}{Sangouard, N.}, \bibinfo{author}{Simon, C.},
  \bibinfo{author}{de~Riedmatten, H.} \& \bibinfo{author}{Gisin, N.}
\newblock \bibinfo{title}{Quantum repeaters based on atomic ensembles and
  linear optics}.
\newblock \emph{\bibinfo{journal}{Rev. Mod. Phys.}}
  \textbf{\bibinfo{volume}{83}}, \bibinfo{pages}{33--80}
  (\bibinfo{year}{2011}).

\bibitem{Chou07}
\bibinfo{author}{Chou, C.W.} \emph{et~al.}
\newblock \bibinfo{title}{Functional quantum nodes for entanglement
  distribution over scalable quantum networks}.
\newblock \emph{\bibinfo{journal}{Science}} \textbf{\bibinfo{volume}{316}},
  \bibinfo{pages}{1316--1320} (\bibinfo{year}{2007}).

\bibitem{Yuan08}
\bibinfo{author}{Yuan, Z.S.} \emph{et~al.}
\newblock \bibinfo{title}{Experimental demonstration of a bdcz quantum repeater
  node}.
\newblock \emph{\bibinfo{journal}{Nature}} \textbf{\bibinfo{volume}{454}},
  \bibinfo{pages}{1098--1101} (\bibinfo{year}{2008}).

\bibitem{Childress06}
\bibinfo{author}{Childress, L.}, \bibinfo{author}{Taylor, J.M.},
  \bibinfo{author}{S\o{}rensen, A.S.} \& \bibinfo{author}{Lukin, M.D.}
\newblock \bibinfo{title}{Fault-tolerant quantum communication based on
  solid-state photon emitters}.
\newblock \emph{\bibinfo{journal}{Phys. Rev. Lett.}}
  \textbf{\bibinfo{volume}{96}}, \bibinfo{pages}{070504}
  (\bibinfo{year}{2006}).

\bibitem{Maiwald09}
\bibinfo{author}{Maiwald, R.} \emph{et~al.}
\newblock \bibinfo{title}{Stylus ion trap for enhanced access and sensing}.
\newblock \emph{\bibinfo{journal}{Nature Phys.}} \textbf{\bibinfo{volume}{5}},
  \bibinfo{pages}{551--554} (\bibinfo{year}{2009}).

\bibitem{Shu10}
\bibinfo{author}{Shu, G.}, \bibinfo{author}{Kurz, N.},
  \bibinfo{author}{Dietrich, M.R.} \& \bibinfo{author}{Blinov, B.B.}
\newblock \bibinfo{title}{Efficient fluorescence collection from trapped ions
  with an integrated spherical mirror}.
\newblock \emph{\bibinfo{journal}{Phys. Rev. A}} \textbf{\bibinfo{volume}{81}},
  \bibinfo{pages}{042321} (\bibinfo{year}{2010}).

\bibitem{Streed11}
\bibinfo{author}{Streed, E.W.}, \bibinfo{author}{Norton, B.G.},
  \bibinfo{author}{Jechow, A.}, \bibinfo{author}{Weinhold, T.J.} \&
  \bibinfo{author}{Kielpinski, D.}
\newblock \bibinfo{title}{Imaging of trapped ions with a microfabricated optic
  for quantum information processing}.
\newblock \emph{\bibinfo{journal}{Phys. Rev. Lett.}}
  \textbf{\bibinfo{volume}{106}}, \bibinfo{pages}{010502}
  (\bibinfo{year}{2011}).

\bibitem{Brady11}
\bibinfo{author}{Brady, G.} \emph{et~al.}
\newblock \bibinfo{title}{Integration of fluorescence collection optics with a
  microfabricated surface electrode ion trap}.
\newblock \emph{\bibinfo{journal}{Appl. Phys. B}}
  \textbf{\bibinfo{volume}{103}}, \bibinfo{pages}{801--808}
  (\bibinfo{year}{2011}).

\bibitem{Merrill11}
\bibinfo{author}{Merrill, J.T.} \emph{et~al.}
\newblock \bibinfo{title}{Demonstration of integrated microscale optics in
  surface-electrode ion traps}.
\newblock \emph{\bibinfo{journal}{New Journal of Physics}}
  \textbf{\bibinfo{volume}{13}}, \bibinfo{pages}{103005}
  (\bibinfo{year}{2011}).

\bibitem{VanDevender10}
\bibinfo{author}{VanDevender, A.P.}, \bibinfo{author}{Colombe, Y.},
  \bibinfo{author}{Amini, J.}, \bibinfo{author}{Leibfried, D.} \&
  \bibinfo{author}{Wineland, D.J.}
\newblock \bibinfo{title}{Efficient fiber optic detection of trapped ion
  fluorescence}.
\newblock \emph{\bibinfo{journal}{Phys. Rev. Lett.}}
  \textbf{\bibinfo{volume}{105}}, \bibinfo{pages}{023001}
  (\bibinfo{year}{2010}).

\bibitem{Hadden10}
\bibinfo{author}{Hadden, J.P.} \emph{et~al.}
\newblock \bibinfo{title}{Strongly enhanced photon collection from diamond
  defect centers under microfabricated integrated solid immersion lenses}.
\newblock \emph{\bibinfo{journal}{Applied Physics Letters}}
  \textbf{\bibinfo{volume}{97}}, \bibinfo{pages}{--} (\bibinfo{year}{2010}).

\bibitem{Robledo11}
\bibinfo{author}{Robledo, L.} \emph{et~al.}
\newblock \bibinfo{title}{High-fidelity projective read-out of a solid-state
  spin quantum register}.
\newblock \emph{\bibinfo{journal}{Nature}} \textbf{\bibinfo{volume}{477}},
  \bibinfo{pages}{574--578} (\bibinfo{year}{2011}).

\bibitem{Casabone13}
\bibinfo{author}{Casabone, B.} \emph{et~al.}
\newblock \bibinfo{title}{Heralded entanglement of two ions in an optical
  cavity}.
\newblock \emph{\bibinfo{journal}{Phys. Rev. Lett.}}
  \textbf{\bibinfo{volume}{111}}, \bibinfo{pages}{100505}
  (\bibinfo{year}{2013}).

\bibitem{Zhang09}
\bibinfo{author}{Zhang, R.}, \bibinfo{author}{Garner, S.R.} \&
  \bibinfo{author}{Hau, L.V.}
\newblock \bibinfo{title}{Creation of long-term coherent optical memory via
  controlled nonlinear interactions in {B}ose-{E}instein condensates}.
\newblock \emph{\bibinfo{journal}{Phys. Rev. Lett.}}
  \textbf{\bibinfo{volume}{103}}, \bibinfo{pages}{233602}
  (\bibinfo{year}{2009}).

\bibitem{Schnorrberger09}
\bibinfo{author}{Schnorrberger, U.} \emph{et~al.}
\newblock \bibinfo{title}{Electromagnetically induced transparency and light
  storage in an atomic {M}ott insulator}.
\newblock \emph{\bibinfo{journal}{Phys. Rev. Lett.}}
  \textbf{\bibinfo{volume}{103}}, \bibinfo{pages}{033003}
  (\bibinfo{year}{2009}).

\bibitem{Schmidt05}
\bibinfo{author}{Schmidt, P.O.} \emph{et~al.}
\newblock \bibinfo{title}{Spectroscopy using quantum logic}.
\newblock \emph{\bibinfo{journal}{Science}} \textbf{\bibinfo{volume}{309}},
  \bibinfo{pages}{749--752} (\bibinfo{year}{2005}).

\bibitem{Maurer12}
\bibinfo{author}{Maurer, P.C.} \emph{et~al.}
\newblock \bibinfo{title}{Room-temperature quantum bit memory exceeding one
  second}.
\newblock \emph{\bibinfo{journal}{Science}} \textbf{\bibinfo{volume}{336}},
  \bibinfo{pages}{1283--1286} (\bibinfo{year}{2012}).

\bibitem{Lidar03}
\bibinfo{author}{Lidar, D.A.} \& \bibinfo{author}{Birgitta~Whaley, K.}
\newblock \bibinfo{title}{Decoherence-free subspaces and subsystems}.
\newblock In \bibinfo{editor}{Benatti, F.} \& \bibinfo{editor}{Floreanini, R.}
  (eds.) \emph{\bibinfo{booktitle}{Irreversible Quantum Dynamics}}, vol.
  \bibinfo{volume}{622} of \emph{\bibinfo{series}{Lecture Notes in Physics}},
  \bibinfo{pages}{83--120} (\bibinfo{publisher}{Springer Berlin Heidelberg},
  \bibinfo{year}{2003}).

\bibitem{Radnaev10}
\bibinfo{author}{Radnaev, A.} \emph{et~al.}
\newblock \bibinfo{title}{A quantum memory with telecom-wavelength conversion}.
\newblock \emph{\bibinfo{journal}{Nat. Phys.}} \textbf{\bibinfo{volume}{6}},
  \bibinfo{pages}{894--899} (\bibinfo{year}{2010}).

\bibitem{Devoret13}
\bibinfo{author}{Devoret, M.H.} \& \bibinfo{author}{Schoelkopf, R.J.}
\newblock \bibinfo{title}{Superconducting circuits for quantum information: An
  outlook}.
\newblock \emph{\bibinfo{journal}{Science}} \textbf{\bibinfo{volume}{339}},
  \bibinfo{pages}{1169--1174} (\bibinfo{year}{2013}).

\bibitem{Lang13}
\bibinfo{author}{Lang, C.} \emph{et~al.}
\newblock \bibinfo{title}{Correlations, indistinguishability and entanglement
  in {H}ong-{O}u-{M}andel experiments at microwave frequencies}.
\newblock \emph{\bibinfo{journal}{Nat. Phys.}} \textbf{\bibinfo{volume}{9}},
  \bibinfo{pages}{345--348} (\bibinfo{year}{2013}).

\bibitem{Kubo10}
\bibinfo{author}{Kubo, Y.} \emph{et~al.}
\newblock \bibinfo{title}{Strong coupling of a spin ensemble to a
  superconducting resonator}.
\newblock \emph{\bibinfo{journal}{Phys. Rev. Lett.}}
  \textbf{\bibinfo{volume}{105}}, \bibinfo{pages}{140502}
  (\bibinfo{year}{2010}).

\bibitem{Schuster10}
\bibinfo{author}{Schuster, D.I.} \emph{et~al.}
\newblock \bibinfo{title}{High-cooperativity coupling of electron-spin
  ensembles to superconducting cavities}.
\newblock \emph{\bibinfo{journal}{Phys. Rev. Lett.}}
  \textbf{\bibinfo{volume}{105}}, \bibinfo{pages}{140501}
  (\bibinfo{year}{2010}).

\bibitem{Zhu11}
\bibinfo{author}{Zhu, X.} \emph{et~al.}
\newblock \bibinfo{title}{Coherent coupling of a superconducting flux qubit to
  an electron spin ensemble in diamond}.
\newblock \emph{\bibinfo{journal}{Nature}} \textbf{\bibinfo{volume}{478}},
  \bibinfo{pages}{221--224} (\bibinfo{year}{2011}).

\bibitem{Amsuess11}
\bibinfo{author}{Ams\"uss, R.} \emph{et~al.}
\newblock \bibinfo{title}{Cavity {QED} with magnetically coupled collective
  spin states}.
\newblock \emph{\bibinfo{journal}{Phys. Rev. Lett.}}
  \textbf{\bibinfo{volume}{107}}, \bibinfo{pages}{060502}
  (\bibinfo{year}{2011}).

\bibitem{Probst13}
\bibinfo{author}{Probst, S.} \emph{et~al.}
\newblock \bibinfo{title}{Anisotropic rare-earth spin ensemble strongly coupled
  to a superconducting resonator}.
\newblock \emph{\bibinfo{journal}{Phys. Rev. Lett.}}
  \textbf{\bibinfo{volume}{110}}, \bibinfo{pages}{157001}
  (\bibinfo{year}{2013}).

\bibitem{Longdell05}
\bibinfo{author}{Longdell, J.J.}, \bibinfo{author}{Fraval, E.},
  \bibinfo{author}{Sellars, M.J.} \& \bibinfo{author}{Manson, N.B.}
\newblock \bibinfo{title}{Stopped light with storage times greater than one
  second using electromagnetically induced transparency in a solid}.
\newblock \emph{\bibinfo{journal}{Phys. Rev. Lett.}}
  \textbf{\bibinfo{volume}{95}}, \bibinfo{pages}{063601}
  (\bibinfo{year}{2005}).

\bibitem{Hedges10}
\bibinfo{author}{Hedges, M.P.}, \bibinfo{author}{Longdell, J.J.},
  \bibinfo{author}{Li, Y.} \& \bibinfo{author}{Sellars, M.J.}
\newblock \bibinfo{title}{Efficient quantum memory for light}.
\newblock \emph{\bibinfo{journal}{Nature}} \textbf{\bibinfo{volume}{465}},
  \bibinfo{pages}{1052--1056} (\bibinfo{year}{2010}).

\bibitem{Simon07}
\bibinfo{author}{Simon, C.} \emph{et~al.}
\newblock \bibinfo{title}{Quantum repeaters with photon pair sources and
  multimode memories}.
\newblock \emph{\bibinfo{journal}{Phys. Rev. Lett.}}
  \textbf{\bibinfo{volume}{98}}, \bibinfo{pages}{190503}
  (\bibinfo{year}{2007}).

\bibitem{Clausen12}
\bibinfo{author}{Clausen, C.}, \bibinfo{author}{Bussi\`eres, F.},
  \bibinfo{author}{Afzelius, M.} \& \bibinfo{author}{Gisin, N.}
\newblock \bibinfo{title}{Quantum storage of heralded polarization qubits in
  birefringent and anisotropically absorbing materials}.
\newblock \emph{\bibinfo{journal}{Phys. Rev. Lett.}}
  \textbf{\bibinfo{volume}{108}}, \bibinfo{pages}{190503}
  (\bibinfo{year}{2012}).

\bibitem{Guendogan12}
\bibinfo{author}{G\"undo\u{g}an, M.}, \bibinfo{author}{Ledingham, P.M.},
  \bibinfo{author}{Almasi, A.}, \bibinfo{author}{Cristiani, M.} \&
  \bibinfo{author}{de~Riedmatten, H.}
\newblock \bibinfo{title}{Quantum storage of a photonic polarization qubit in a
  solid}.
\newblock \emph{\bibinfo{journal}{Phys. Rev. Lett.}}
  \textbf{\bibinfo{volume}{108}}, \bibinfo{pages}{190504}
  (\bibinfo{year}{2012}).

\bibitem{Zhou12}
\bibinfo{author}{Zhou, Z.Q.}, \bibinfo{author}{Lin, W.B.},
  \bibinfo{author}{Yang, M.}, \bibinfo{author}{Li, C.F.} \&
  \bibinfo{author}{Guo, G.C.}
\newblock \bibinfo{title}{Realization of reliable solid-state quantum memory
  for photonic polarization qubit}.
\newblock \emph{\bibinfo{journal}{Phys. Rev. Lett.}}
  \textbf{\bibinfo{volume}{108}}, \bibinfo{pages}{190505}
  (\bibinfo{year}{2012}).

\bibitem{Chen13}
\bibinfo{author}{Chen, W.} \emph{et~al.}
\newblock \bibinfo{title}{All-optical switch and transistor gated by one stored
  photon}.
\newblock \emph{\bibinfo{journal}{Science}} \textbf{\bibinfo{volume}{341}},
  \bibinfo{pages}{768--770} (\bibinfo{year}{2013}).

\end{thebibliography}

\section{Acknowledgments}

\begin{acknowledgments}
We thank S. Olmschenk, C. Monroe, W. Rosenfeld, J. Majer, P. Bushev, and H. de Riedmatten for providing images and feedback
and G. Kirchmair, S. Ritter, B. Casabone, K. Friebe, and Y. Colombe for helpful comments.
We gratefully acknowledge support from the Austrian Science Fund (FWF):  Project Nos. F4002-N16 and F4019-N16, 
the European Research Council through the CRYogenic Traps for
Entanglement Research with IONs (CRYTERION) Project,
the European Commission via the Atomic QUantum TEchnologies (AQUTE) Integrating Project, 
the Intelligence Advanced Research Projects Agency,
and the Institut f\"ur Quanteninformation GmbH.
\end{acknowledgments}

\end{document}